\documentclass[apjl]{emulateapj}
\usepackage{amsfonts,amsmath,graphicx,natbib,apjfonts,subfigure,color,gensymb,longtable}

\def\ra#1#2#3{#1$^{\rm h}$#2$^{\rm m}$#3$^{\rm s}$}
\def\dec#1#2#3{#1$^\circ$#2$'$#3$''$}

\def\gw{GW170817A}
\def\nuu{1}
\def\cfa{2}
\def\hf{3}
\def\fr{4}
\def\columbia{5}
\def\oh{6}
\def\fermi{7}
\def\penn{12}
\def\si{8}
\def\kavli{10}
\def\sc{14}
\def\hst{15}
\def\kavlidan{11}
\def\brandeis{13}
\def\chi{9}
\shorttitle{\gw X-rays}
\shortauthors{Margutti~et~al.}

\begin{document}
\title{The Electromagnetic Counterpart of the Binary Neutron Star Merger LIGO/VIRGO GW170817. V. Rising X-ray Emission from an Off-Axis Jet}

\author{R.~Margutti\altaffilmark{\nuu}, 
E.~Berger\altaffilmark{\cfa}, 
W.~Fong\altaffilmark{\nuu}$^,$\altaffilmark{\hf},
C.~Guidorzi\altaffilmark{\fr},
K.~D.~Alexander\altaffilmark{\cfa},
B.~D.~Metzger\altaffilmark{\columbia},
P.~K.~Blanchard\altaffilmark{\cfa},
P.~S.~Cowperthwaite\altaffilmark{\cfa},
R.~Chornock\altaffilmark{\oh},
T.~Eftekhari\altaffilmark{\cfa},
M.~Nicholl\altaffilmark{\cfa},
V.~A.~Villar\altaffilmark{\cfa},
P.~K.~G.~Williams\altaffilmark{\cfa},
J.~Annis\altaffilmark{\fermi},
%D. Brout\altaffilmark{\penn},
D.~A. Brown\altaffilmark{\si},
H.~Y. Chen\altaffilmark{\chi},
Z. Doctor\altaffilmark{\chi},
%T. Diehl\altaffilmark{\fermi},  OPT OUT
%R. Foley\altaffilmark{\sc},
J.~A.~Frieman\altaffilmark{\fermi,\kavli},
%K. Herner\altaffilmark{\fermi},
D.~E.~Holz\altaffilmark{\kavlidan,\kavli},
%R. Kessler\altaffilmark{\kavli},
%A. Rest\altaffilmark{\hst},
M. Sako\altaffilmark{\penn},
M. Soares-Santos\altaffilmark{\brandeis,\fermi}
}

\altaffiltext{\nuu}{Center for Interdisciplinary Exploration and Research in Astrophysics (CIERA) and Department of Physics and Astronomy, Northwestern University, Evanston, IL 60208}
\altaffiltext{\cfa}{Harvard-Smithsonian Center for Astrophysics, 60 Garden Street, Cambridge, MA 02138, USA}
\altaffiltext{\hf}{Hubble Fellow}
\altaffiltext{\fr}{Department of Physics and Earth Science, University of Ferrara, via Saragat 1, I--44122, Ferrara, Italy}
\altaffiltext{\columbia}{Department of Physics and Columbia Astrophysics Laboratory, Columbia University, New York, NY 10027, USA}
\altaffiltext{\oh}{Astrophysical Institute, Department of Physics and Astronomy, 251B Clippinger Lab, Ohio University, Athens, OH 45701, USA}
\altaffiltext{\fermi}{Fermi National Accelerator Laboratory, P.O. Box 500, Batavia, IL 60510, USA}
\altaffiltext{\si}{Physics Department, Syracuse University, Syracuse, NY 13244, USA}
\altaffiltext{\chi}{Department of Astronomy and Astrophysics, University of Chicago, Chicago, Illinois 60637, USA}
\altaffiltext{\kavli}{Kavli Institute for Cosmological Physics, University of Chicago, Chicago, IL 60637, USA}
\altaffiltext{\kavlidan}{Enrico Fermi Institute, Department of Physics, Department of Astronomy and Astrophysics, University of Chicago, Chicago, IL 60637, USA}
\altaffiltext{\penn}{Department of Physics and Astronomy, University of Pennsylvania, Philadelphia, PA 19104, USA}
\altaffiltext{\brandeis}{Department of Physics, Brandeis University, Waltham, MA 02454, USA}
\altaffiltext{\sc}{Department of Astronomy and Astrophysics, University of California, Santa Cruz, CA 95064, USA}
\altaffiltext{\hst}{Department of Physics and Astronomy, The Johns Hopkins University, 3400 North Charles Street, Baltimore, MD 21218, USA}

\begin{abstract}
We report the discovery of rising X-ray emission from the binary neutron star (BNS) merger event GW170817. This is the first detection of X-ray emission from a gravitational-wave source. Observations acquired with the {\it Chandra X-ray Observatory} (CXO) at $t\approx 2.3$ days post merger reveal no significant emission, with $L_x\lesssim 3.2\times 10^{38}\,\rm{erg\,s^{-1}}$ (isotropic-equivalent). Continued monitoring revealed the presence of an X-ray source that brightened with time, reaching $L_x\approx 9\times 10^{39}\,\rm{erg\,s^{-1}}$ at $\approx 15.1$ days post merger. We interpret these findings in the context of isotropic and collimated relativistic outflows (both on- and off-axis). We find that the broad-band X-ray to radio observations are consistent with emission from a relativistic jet with kinetic energy $E_{k}\sim 10^{49-50}\,\rm{erg}$, viewed off-axis with $\theta_{\rm obs}\sim 20-40\degree$. Our models favor a circumbinary density $n\sim 10^{-4}-10^{-2}\,\rm{cm^{-3}}$,  depending on the value of the microphysical parameter $\epsilon_B=10^{-4}-10^{-2}$.  A central-engine origin of the X-ray emission is unlikely. Future X-ray observations at $t\gtrsim 100$ days, when the target will be observable again with the CXO, will provide additional constraints to solve the model degeneracies and test our predictions. Our inferences on $\theta_{obs}$ are testable with gravitational wave information on GW170817 from Advanced LIGO/Virgo on the binary inclination.
\end{abstract}
%%%%%%%%%%%%%%%%%%%%%%%%%%%%%%%%%%%%%%%%%%%
\section{Introduction}
\label{Sec:intro}
Gravitational waves (GW) from the merger of a binary neutron star (BNS) system were detected for the first time by 
Advanced LIGO and Advanced Virgo on 2017 August 17.53 UT \citep{ALVgcn,ALVdetection}. The GW event, named GW170817, was localized to a region of $\sim 30$ deg$^2$ with a distance of $\sim40$ Mpc. The GW signal from the BNS merger was closely followed in time by a short burst of $\gamma$-ray emission detected by \emph{Fermi} and Integral (\citealt{GBMgcn1,INTEGRALgcn}) with fluence $F_{\gamma}=(2.4\pm 0.5)\times 10^{-7}\,\rm{erg\,cm^{-2}}$ \citep{GBMdetection}. These observations established GW170817 to be the first astrophysical event with GW and EM detections, marking the dawn of multi-messenger astrophysics.\footnote{Note that the optical transient source was given the name of SSS17a (\citealt{SWOPEgcn,SWOPEpaper}) and DLT17ck (\citealt{DLT40gcn,ValentiPaper}), as well as an International Astronomical Union name of AT2017gfo.}

Optical observations acquired within $\sim 12$ hours after the GW detection led to the discovery and localization of a transient with peculiar properties in the outskirts of the galaxy NGC\,4993 %by our team and others
\citep{SWOPEgcn,DECAMgcn,DLT40gcn,DECamPaper1,PeterPaper,ValentiPaper}; see Soares-Santos et al. 2017 for details of our group's discovery. Intense photometric and spectroscopic UV/optical/NIR monitoring of the transient (\citealt{PhilPaper,MattPaper,RyanPaper}) revealed an evolution that closely follows the theoretical expectations from  a ``kilonova'' (KN),  i.e. a transient powered by the radioactive decay of $r$-process nuclei synthesized in the neutron-rich merger ejecta (see \citealt{Metzger17} for a recent review). 

Non-thermal radiation at X-ray and radio wavelengths is also expected to be associated with BNS mergers on different timescales and luminosities \emph{if} these systems are able to launch relativistic jets, as initially postulated in the case of Short Gamma-Ray Bursts (SGRB, e.g. \citealt{Eichler89,Narayan92}). Observations of the environments and properties of emission of SGRBs in the last decade provided solid indirect evidence of the association of SGRBs with binary NS mergers \citep{Fong13,Berger14,Fong15}, thus motivating our search for observational signatures of on-axis and off-axis jets in GW170817.

Here, we report the first detection of X-ray emission from a GW event. We explore various scenarios for the origin of the X-ray emission, and place constraints on the properties of the circumbinary medium, jet energetics, collimation and observer angle based on the broad-band X-ray to radio observations. A comparison to the properties of ``canonical" SGRBs can be found in \cite{WFpaper}, while we refer to our companion paper \cite{KatePaper} for a dedicated discussion of the radio observations of GW170817. Our X-ray observations of NGC\,4993, the  host galaxy of GW170817, are discussed in \cite{PeterPaper}.

We assume a distance to NGC\,4993 of 39.5\,Mpc ($z=0.00973$) as listed in the NASA Extragalactic Database. $1\,\sigma$ c.l. uncertainties are listed unless otherwise stated. In this manuscript we employ the notation $Q_x\equiv Q/10^x$. In this paper we always refer to isotropic-equivalent luminosities. We differentiate between isotropic-equivalent kinetic energy $E_{k,iso}$, and beaming-corrected kinetic energy of the blast wave $E_{k}$, where $E_{k}=E_{k,iso}(1-cos(\theta_j))$ and $\theta_j$ is the jet opening angle. 
%%%%%%%%%%%%%%%%%%%%%%%%%%%%%%%%%%%%%%%%%%
\section{Observations}
\label{Sec:Obs}

\begin{figure*}
\center
\includegraphics[scale=0.5]{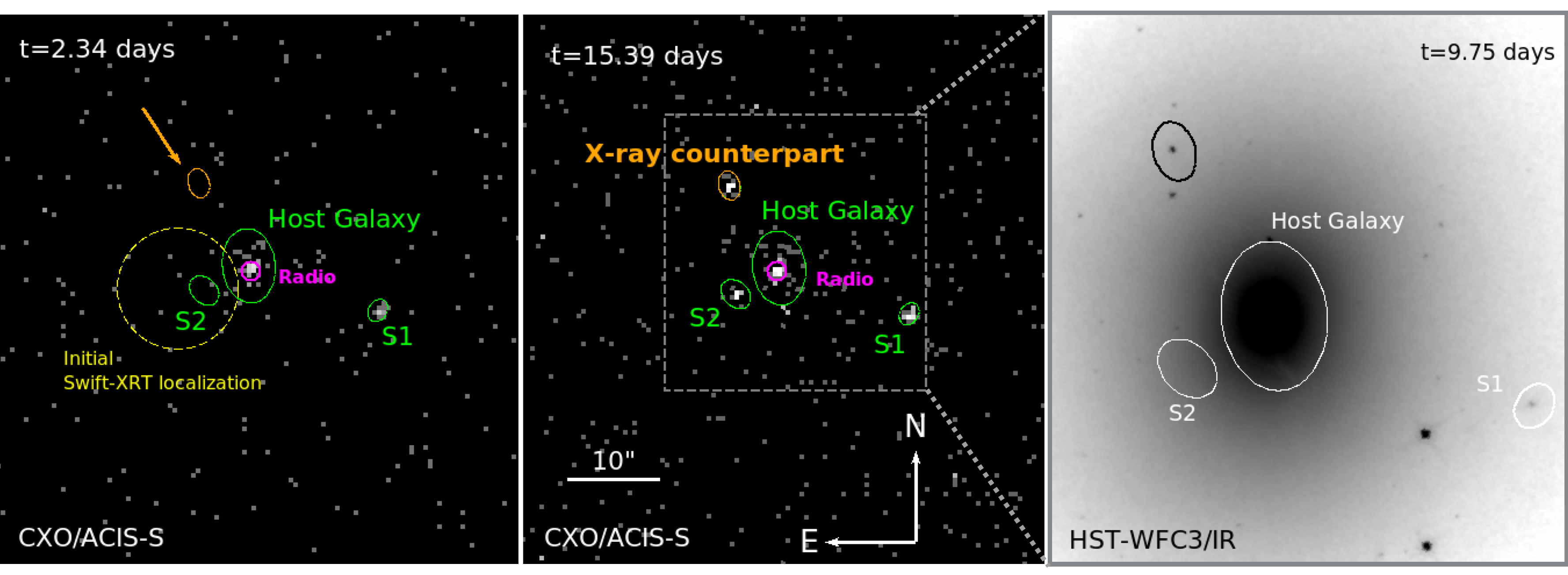}
\caption{0.5-8 keV CXO observations of the optical transient associated with GW170817 obtained at $\sim$2.34 days (left panel) and $\sim$15.39 days (central panel) since BNS coalescence reveal the appearance of a new X-ray source at the location of the optical transient (right panel). The host galaxy is a source of diffuse and persistent X-ray emission, with the core of the X-ray emission coincident with the radio source (1" magenta region) that we identified in \cite{Kategcn}, suggesting the presence of a weak AGN \citep{PeterPaper}.  The central panel also shows the appearance of another X-ray source S2, which was not detected in our first CXO observations. The initial localization of an X-ray source by the \emph{Swift}-XRT at $t<2$ days \citep{SWIFTgcn1} (yellow dashed region in the left panel, 90\% containment) might suggest that S2 was ``active" before our first CXO  observation. Right panel: zoom-in into HST observations of the EM counterpart to GW170817 \citep{MattPaper,PeterPaper} with the X-ray regions overlaid. }
\label{Fig:ImageChandra}
\end{figure*}

%brief description of localization with DECam? or just refer to papers....

With the Dark Energy Camera, we independently discovered and localized the optical transient to RA=\ra{13}{09}{48.08}, Dec=\dec{-23}{22}{53.2} (J2000) with $1\sigma$ uncertainties of 130~mas and 60~mas, respectively \citep{DECamPaper1}, and initiated multi-wavelength follow up of the transient across the electromagnetic spectrum.  Here we report on X-ray observations that led to the first identification of rising X-ray emission from a binary neutron star merger event GW170817. 

%-------------------------------------------Swift-XRT
\subsection{Swift X-ray Observations}
\label{SubSec:XRT}

The \emph{Swift} spacecraft \citep{Gehrels04} started observations of the optical counterpart of LIGO/Virgo GW170817  (\citealt{SWOPEgcn,SWOPEpaper}; \citealt{DECAMgcn,DECamPaper1}; \citealt{DLT40gcn}) with the X-ray Telescope (XRT, \citealt{Burrows05}) on August 18th, 03:33:33UT, 14.9 hrs after the GW trigger.
Swift-XRT observations span the time range $0.6-11.5$ days since trigger, at which point the target entered into Sun constraint.   Swift-XRT data have been analyzed using HEASOFT (v6.22) and corresponding calibration files, employing standard filtering criteria and following standard procedures (see \citealt{Margutti13} for details). No transient X-ray emission is detected at the location of the GW optical counterpart (\citealt{SWIFTgcn1,SWIFTgcn2,SWIFTpaper}), with typical count-rate limits of $\sim$ a few $10^{-3}\,\rm{c\,s^{-1}}$. The neutral Hydrogen column density in the direction of the transient is $NH_{mw}=0.0784\times 10^{22}\,\rm{cm^{-2}}$ \citep{Kalberla05}. For a typical absorbed power-law spectrum with photon index $\Gamma\sim2$ and negligible intrinsic absorption (see below), the corresponding $3\,\sigma$ flux limit is $\sim10^{-13}\,\rm{erg\,s^{-1}\,cm^{-2}}$ (unabsorbed, 0.3-10 keV), which is $L_x<$  a few $10^{40}\,\rm{erg\,s^{-1}}$ at the distance of 39.5 Mpc. As we show in detail in \cite{WFpaper}, \emph{Swift}-XRT observations constrain the X-ray emission associated with the optical counterpart of LIGO/Virgo GW170817 to be significantly fainter  than cosmological short GRBs at the same epoch (\citealt{Margutti13,Fong15,DAvanzo14}).

%-------------------------------------------CHANDRA
\subsection{Chandra X-ray Observations}
\label{SubSec:Chandra}

We initiated deep X-ray follow up of the optical transient with the \emph{Chandra} X-ray Observatory  (CXO) on 2017 August 19.71UT,  $\delta t\approx 2.3\,\rm{d}$ after the GW detection (observation ID 18955; PI:~Fong; Program 18400052). \emph{Chandra} ACIS-S data have been reduced with the {\tt CIAO} software package
(v4.9) and relative calibration files, applying standard ACIS data filtering. Using {\tt wavdetect} we find no evidence for X-ray emission at the position of the optical transient 
\citep{Chandra1gcn}  and we infer a 
$3\sigma$ limit of $1.2\times 10^{-4}\rm{cps}$ (0.5-8 keV
energy range, total exposure time of $24.6$ ks). For an assumed absorbed spectral power-law model with $\Gamma=2$, negligible intrinsic absorption and $NH_{mw}=0.0784\times 10^{22}\,\rm{cm^{-2}}$, the corresponding absorbed (unabsorbed) flux limit in the 0.3-10 keV energy range is $F_{\rm{x}}<1.4\times 10^{-15}\,\rm{erg\,s^{-1}cm^{-2}}$  ($F_{\rm{x}}<1.7\times 10^{-15}\,\rm{erg\,s^{-1}cm^{-2}}$).\footnote{Significant intrinsic absorption is not expected, given the early-type nature of the host galaxy and the location of the transient in the outskirts of its host galaxy, \citep{PeterPaper}. This expectation is independently confirmed by our optical/NIR modeling \citep{PeterPaper}, which indicates $NH_{int}<10^{21}\,\rm{cm^{-2}}$, and by the X-ray analysis of the epoch when the transient is detected. However, we repeated our analysis of the first CXO epoch focusing on the harder part of the spectrum to minimize the possible effects of absorption. We find a $3\sigma$ limit of $1.2\times 10^{-4}\rm{cps}$ (0.8-8 keV), which corresponds to a limit on the flux density at 1 keV $F_{1keV}<1.40\times 10^{-4}\,\rm{\mu Jy}$. With the previous spectral calibration we would infer a similar value $F_{1keV}<1.32\times 10^{-4}\,\rm{\mu Jy}$. We conclude that our modeling below, which employs  $F_{1keV}$ is  thus robust.} The luminosity limit is  $L_{\rm{x}}<3.2\times 10^{38}\,\rm{erg\,s^{-1}}$ (0.3-10 keV), making the X-ray counterpart to GW170817  $\ge1000$ times fainter than on-axis short GRBs at the same epoch (\citealt{WFpaper}). 

We re-visited the location of the optical transient on September 1.64UT (starting 15.1 days since trigger) under a DDT program with shared data (observation ID 20728; data shared among Troja, Haggard, Margutti; Program 18508587) with an effective exposure time of 46.7 ks. An X-ray source is blindly detected \citep{Chandra2gcn} with high significance of $\sim7.3\,\sigma$ at  RA=\ra{13}{09}{48.076} and  Dec=$-$\dec{23}{22}{53.34} (J2000), see Fig. \ref{Fig:ImageChandra}, consistent with the optical transient and the findings by \cite{TrojaGCN}.

The source 0.5-8 keV count-rate is $(3.8  \pm 0.9)\times 10^{-4}\,\rm{cps}$. The total number of 0.5-8 keV counts in the source region is 19.  Based on Poissonian statistics, the probability to observe 0 events in 24.6 ks (as in our first observation), if the expected rate is 19 events in 46.7 ks, is $\sim 0.0045$\% ($\sim4$ Gaussian $\sigma$ equivalent). A similar result is obtained with a Binomial test ($P\sim 0.03$\%, corresponding to $\sim3.6$ Gaussian $\sigma$). \emph{We can thus reject the hypothesis of a random fluctuation of a persistent X-ray source with high confidence, and we conclude that we detected rising X-ray emission in association to the optical counterpart to GW170817.} 

The limited statistics does not allow us to constrain the spectral model. We employ Cash statistics to fit the spectrum with an absorbed power-law spectral model with index $\Gamma$ and perform a series of MCMC simulations to constrain the spectral parameters. We find $\Gamma=1.6^{+1.5}_{-0.1}$ ($1\,\sigma$ c.l.) with no evidence for intrinsic neutral hydrogen absorption $\rm{NH_{int}}<3 \times 10^{22}\,\rm{cm^{-2}}$ ($3\,\sigma$ c.l.).For these parameters, the inferred 0.3-10 keV flux is $(3.0-5.6)\times 10^{-15}\,\rm{erg\,s^{-1}\, cm^{-2}}$ ($1\,\sigma$ c.l.). The corresponding unabsorbed flux is $(3.1-5.8)\times 10^{-15}\,\rm{erg\,s^{-1}\, cm^{-2}}$, luminosity $L_x$ in the range $(5.9-11.1)\times 10^{38}\,\rm{erg\,s^{-1}}$ ($1\,\sigma$ c.l.). 

Figure \ref{Fig:Off1} shows our CXO light-curve of the X-ray source associated with GW170817. In this figure we add the X-ray measurement by \cite{HaggardGCN1,HaggardPaper} obtained $15.9$ days after GW trigger (PI Haggard, ID 18988) and rescaled to $\Gamma=2$ in the 0.3-10 keV energy range, leading to $F_x\sim4.5\times 10^{-15}\,\rm{erg\,s^{-1}\, cm^{-2}}$. This flux is consistent with our observations obtained $\sim24$ hrs before, with no statistically significant evidence for temporal variability of the source on this timescale. An estimate of the lower limit of the X-ray flux at $t\sim10$ days, corresponding to the reported detection of X-ray emission with the CXO using an exposure time of 50 ks (\citealt{TrojaGCN}) is also shown to guide the eye.

%%%%%%%%%%%%%%%%%%%%%%%%%%%%%%%%%%%%%%%%%%
\section{Origin of the rising X-ray emission}
\label{Sec:origin}
We discuss the physical origin of the rising X-ray emission found in association to GW170817 considering the following observational constraints: (i) The peak of the X-ray emission is at $t_{pk}\ge 15$ days; (ii) The X-ray light-curve shows mild temporal evolution, with no signs of rise or decay over a $\sim$24 hr timescale at $t\sim15$ days; (iii) The blue colors of the early kilonova emission (\citealt{PhilPaper,MattPaper}) suggest $\theta_{obs}<45\degree$ \citep{Sekiguchi16}, where $\theta_{obs}$ is the observer angle with respect to the jet axis (Sec. \ref{SubSec:off});\footnote{As a note of caution, we mention here that it might be possible to observe blue emission from a kilonova even from larger viewing angles if it expands faster than the tidal matter. This scenario has yet to be fully explored.} (iv) Simultaneous radio observations from \cite{KatePaper} which include the earliest radio observations of this transient at different frequencies and detections at  6 GHz.  Below we discuss the nature of the X-ray emission from GW170817 considering this entire range of observational constraints available at the time of writing.

\subsection{Constraints on on-axis outflows}
We first consider constraints on on-axis\footnote{i.e. Outflows for which $\theta_{obs} \le \theta_j$, where $\theta_j$ is the half-opening angle of the core of the jet and $\theta_{obs}$ is the observer angle with respect to the jet axis.} relativistic outflows (collimated or not collimated), under the assumption that the the blast wave has transferred to the  ISM most of its energy by the time of our first CXO observation, and its hydrodynamics is thus well described by the Blandford-McKee (BM) self-similar deceleration solution \citep{Blandford76}. Electrons are accelerated at the shock front into a power-law distribution  $N(\gamma)\propto \gamma^{-p}$ for $\gamma\ge \gamma_{min}$ and cool through synchrotron emission and adiabatic losses. 

%We first explore the scenario that the X-ray emission originates from the afterglow viewed at an angle $\theta_{\rm obs}$ from the jet axis (``off-axis afterglow'') by comparing the observations to models. For the afterglow emission, we utilize the standard synchrotron model for a relativistic blast-wave in a constant density medium \citep{Granot02}, as expected for a non-massive star progenitor. This model provides a mapping from the afterglow brightness and evolution to the isotropic-equivalent kinetic energy ($E_{\rm K,iso}$), circum-merger density ($n$), fractions of post-shock energy in radiating electrons ($\epsilon_e$) and magnetic fields ($\epsilon_B$), and the electron power-law distribution index ($p$), with $N(\gamma)\propto \gamma^{-p}$ for $\gamma \gtrsim \gamma_{\rm min}$, where $\gamma_{\rm min}$ is the minimum Lorentz factor of the electron distribution.

In the standard synchrotron model (e.g. \citealt{Granot02}), the flux density $F_{\nu}\propto n^{1/2}E_{k,iso}^{(3+p)/4}\epsilon_e^{p-1}\epsilon_B^{(1+p)/4}\,t^{(3-3p)/4}$ if the X-rays are on the $\nu^{(1-p)/2}$ spectral segment (i.e. $\nu_x<\nu_c$) and $F_{\nu}\propto E_{k,iso}^{(2+p)/4}\epsilon_e^{p-1}\epsilon_B^{(p-2)/4}\,t^{(2-3p)/4}$ if the X-rays are on the $\nu^{-p/2}$ spectral segment ($\nu_x>\nu_c$). $\nu_c$ is the synchrotron cooling frequency (e.g. \citealt{Rybicki79}), $\epsilon_e$  and $\epsilon_B$ are the post-shock energy fractions in electrons and magnetic field, respectively and $n$ is the ISM density. We use a constant density medium as expected for a non-massive star progenitor. 

Within this model, and for fiducial parameters $\epsilon_e=0.1$, $\epsilon_B=0.01$ and $p=2.4$ set by median value of cosmological short GRBs (\citealt{Fong15}), the deep CXO non-detection on day 2.34 constrains $E_{k,iso}\leq 10^{47}n_0^{-10/27}\,\rm{erg}$ for  $\nu_x<\nu_c$ and $E_{k,iso}\leq4\times 10^{46}\,\rm{erg}$ for $\nu_x>\nu_c$. $n_0$ is the circumburst density in units of $\rm{cm^{-3}}$. Consistent with the results from radio observations \citep{KatePaper}, this analysis points at low $E_{k,iso}\leq 10^{48}\,\rm{erg}$ for the range of densities $n\sim(3-15)\times 10^{-3}\,\rm{cm^{-3}}$ associated to cosmological short GRBs, which are characterized by $E_{k,iso}\sim (1-3)\times 10^{51}\,\rm{erg}$ for the same microphysical parameters $\epsilon_e=0.1$ and $\epsilon_B=0.01$ \citep{Fong15}. We note that this conclusion does not depend on the choice of $p$, with  $p=2.1-2.4$ ($p>2.4$ violates our radio limits). This solution is only valid during the relativistic phase at $t<t_{NR}$ (where $t_{NR}\sim 1100 \,(E_{k,iso}/10^{53} n_0)^{1/3}\,\rm{days}$, \citealt{Piran04}) and constraints the presence of an undetected, temporally \emph{decaying} X-ray emission at $t<2.34$ days, with properties that are clearly distinguished from cosmological short GRBs seen on-axis \citep{WFpaper}. 

A \emph{rising} X-ray light-curve can be the result of a delayed onset of the afterglow emission, as the blast wave decelerates into the environment and transfers energy to the circumburst medium. In this scenario,  the initial Lorentz factor of the outflow is $\Gamma_{0}\sim 8.0\,E_{k,iso,52}^{1/8}n_{0}^{-1/8}t_{pk,day}^{-3/8}$ where $t_{pk,day}$ is the peak time of the afterglow in days \citep{Sari99b}. A distinguishing feature of the early afterglow emission is an initial very steep rise of the emission $\propto t^2$ or $\propto t^{11/3}$ \citep{Sari99b}. The stable X-ray flux of the source at $t\sim15-16$ days suggests that $t_{pk}\sim 15-30$ days. Given the Fermi-GBM detection of a gamma-ray transient with fluence $F\sim2.4\times 10^{-7}\,\rm{erg\,cm^{-2}}$ \citep{GBMdetection}, which gives $E_{k,iso}\sim5\times 10^{47}\,\rm{erg}$ for a fiducial $\gamma$-ray efficiency $\eta_{\gamma}=0.1$, we infer  a mildly relativistic $\Gamma_{0}\sim 2$ for $t_{pk}\sim 15-30$ days. 
After peak, when most of the fireball energy has been transferred to the ISM, the standard afterglow scalings apply. The latest CXO detection implies $E_{k,iso}\sim 10^{48}n_0^{-10/27}\,\rm{erg}$ if $\nu_x<\nu_c$, or $E_{k,iso}\sim 10^{48}\,\rm{erg}$ for $\nu_x>\nu_c$. %The requirement $\nu_x>\nu_c$ further imposes an additional constraint in the ($E_{k,iso}, n$) parameter space as $\nu_c\propto n_0^{-1}E_{k,iso}^{-1/2}$ (\citealt{Granot02}). 
Radio observations acquired around the same time \citep{KatePaper} constrain $p\approx2.2$. Mildly-relativistic outflows with similar $\Gamma$ and $E_{k}$ that are found in shocks from supernovae (SN) with fast ejecta (i.e. relativistic SNe)  are well described by $p\sim3$ (e.g. \citealt{Chevalier06,Soderberg10,Chakraborti15}). From a purely theoretical perspective, both analytical models and PIC (particle-in-cell) simulations confirm that $p=2.2$ is expected in the cases of \emph{ultra-relativistic shocks where particle acceleration is very efficient.} We thus conclude that a late onset of a weak on-axis afterglow emission is unlikely to provide a satisfactory explanation of our observations across the electromagnetic spectrum, and we consider alternative explanations below. 

%%%%%%%%%%%%%%%%%%%%%%%%%%%%%%%%%%%%%%%%%%
\subsection{Constraints on Off-axis Jets}
\label{SubSec:off}

\begin{figure*}[t!]
\center
 \includegraphics[scale=0.47]{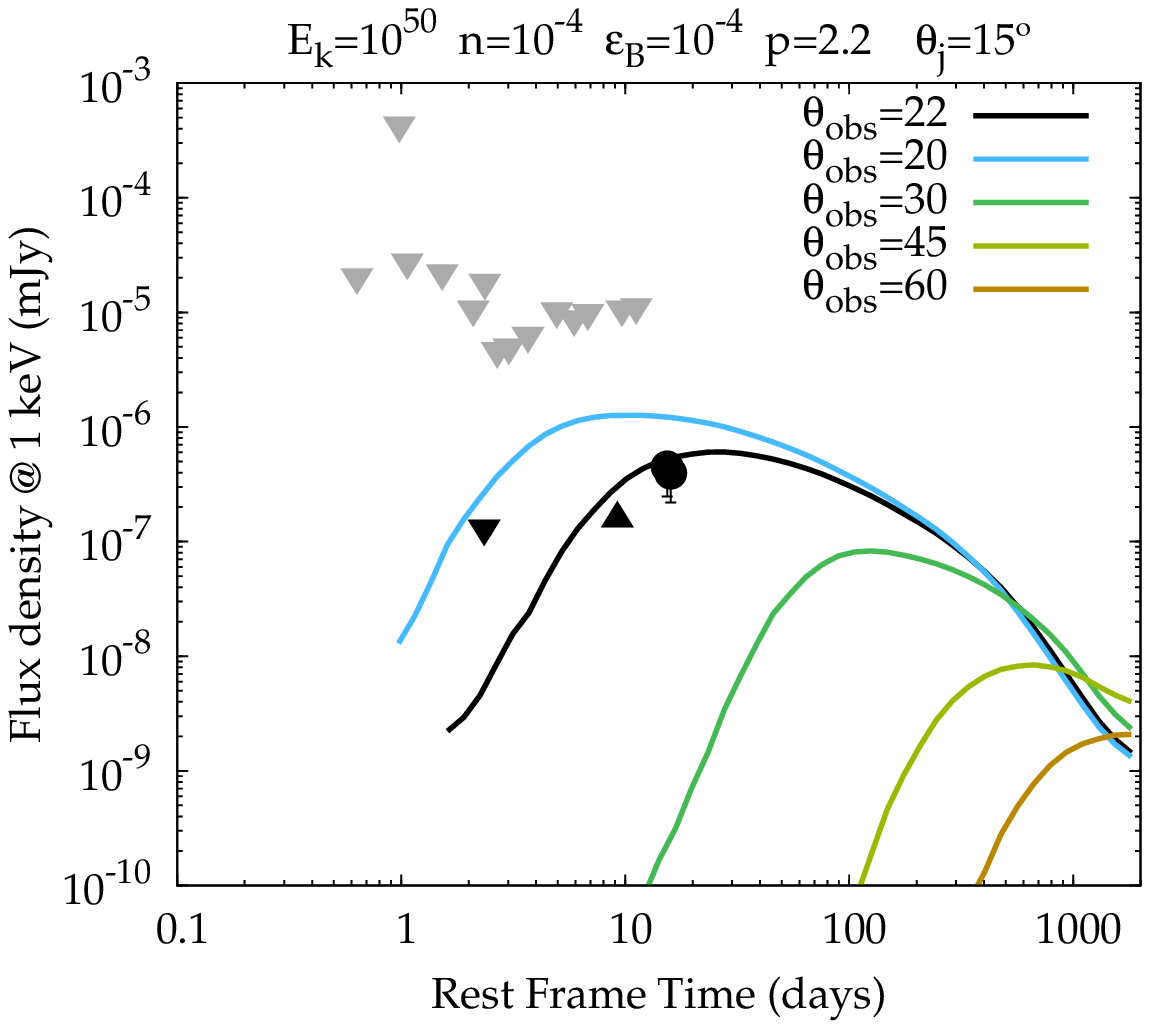}
 \includegraphics[scale=0.47]{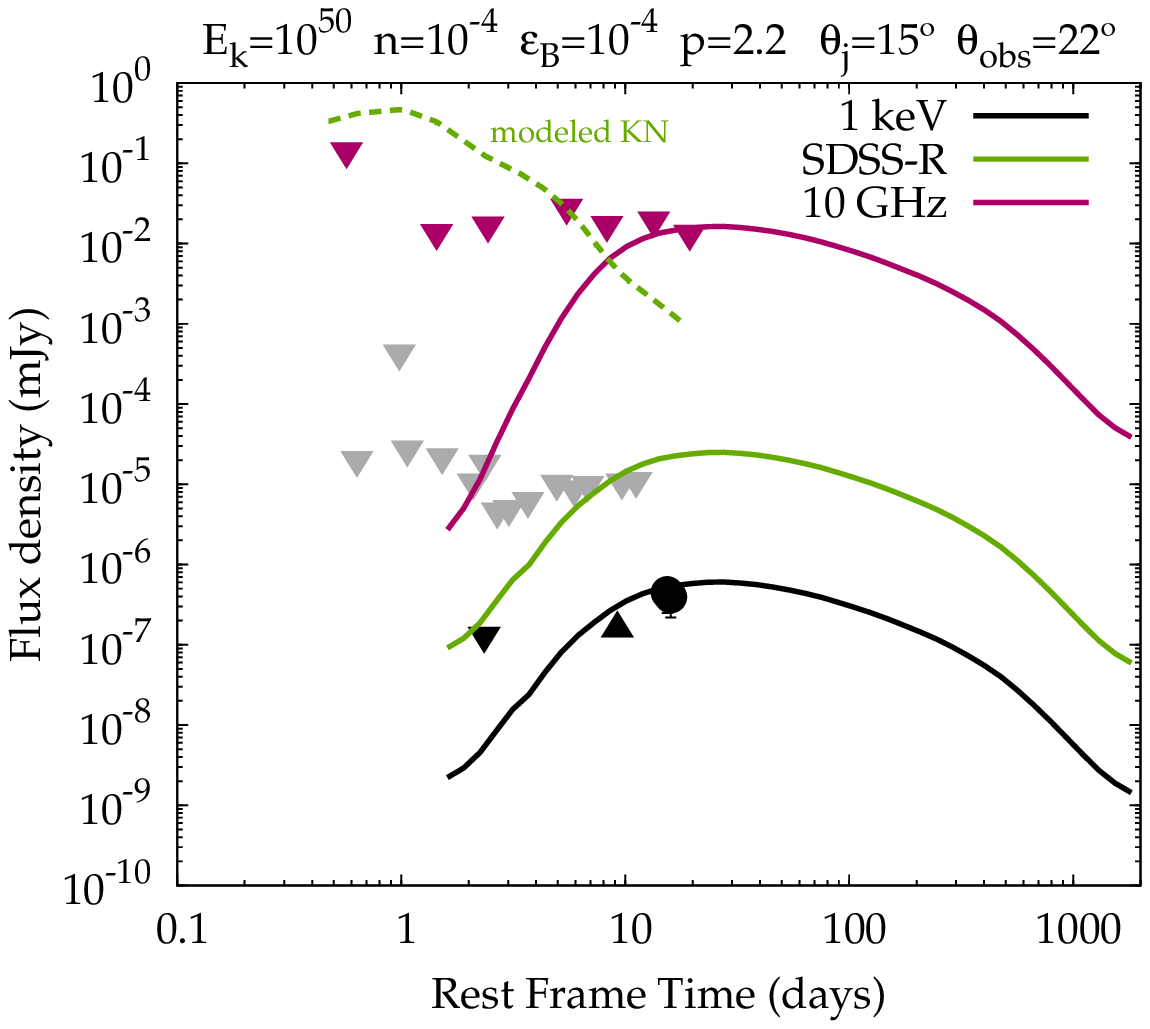}
 \includegraphics[scale=0.45]{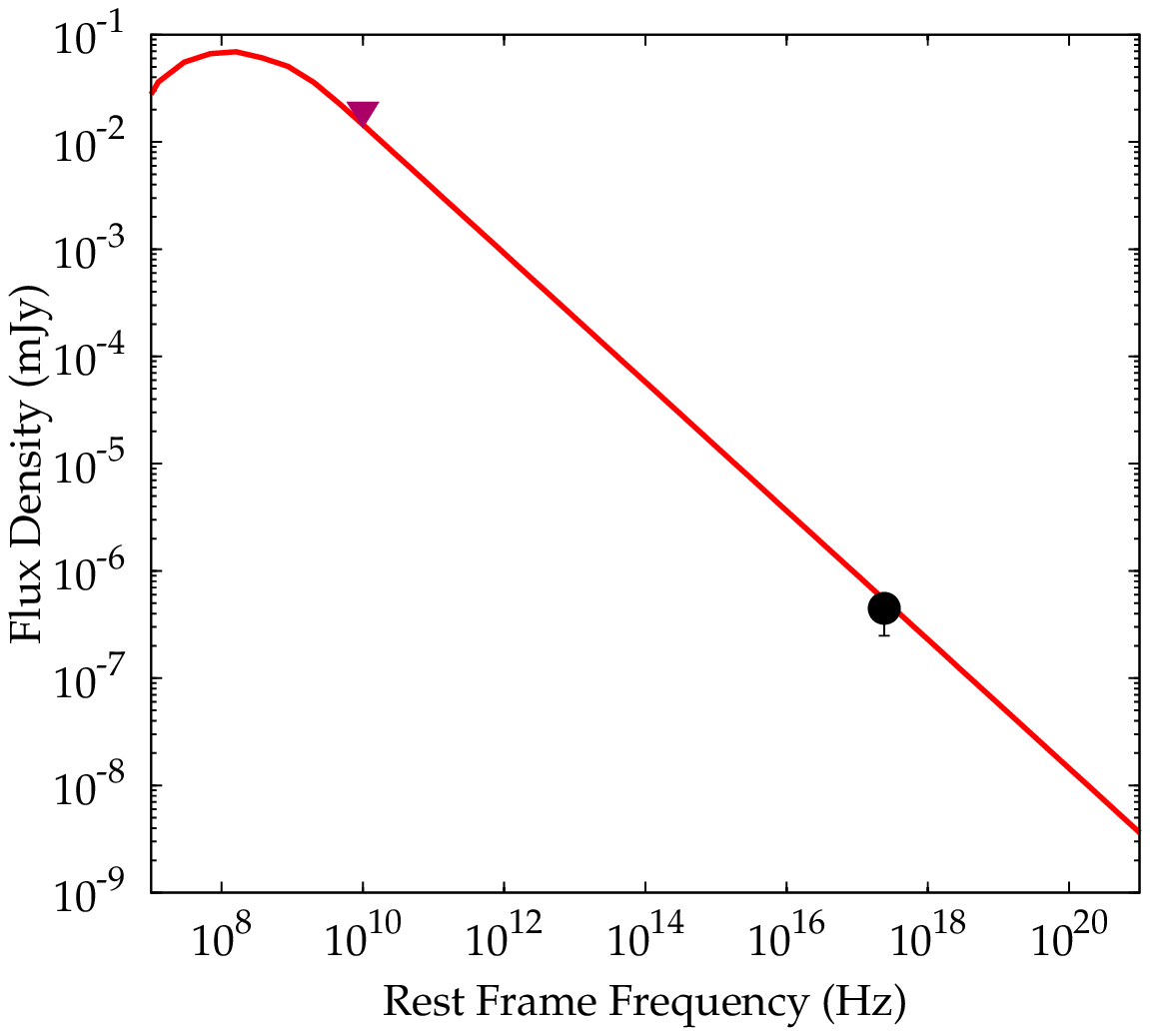}\\
\caption{Off-axis jet model with $\theta_j=15\degree$ and $E_{k}=10^{50}\,\rm{erg}$ that best represents the current set of X-ray and radio observations (see Fig. \ref{Fig:Off2} for models with $E_{k}=10^{49}\,\rm{erg}$). For this model, $n=10^{-4}\,\rm{cm^{-3}}$, $\epsilon_B=10^{-4}$.   \emph{Left panel}: X-ray emission for observers at different $\theta_{obs}$ (colored lines). The black line identifies the best-fitting model, which has $\theta_{obs}\sim22\degree$. Grey triangles: \emph{Swift}-XRT upper limits. Black symbols: CXO observations. We show the results from \cite{TrojaGCN} as an upward triangle (lower limit) for graphics purposes only.  \emph{Central panel}: radio (10 GHz, solid purple line) and optical emission (r-band, solid green line) for the best-fitting model compared to our VLA limits (purple triangles, \citealt{KatePaper}) and emission from the kilonova (green dashed line, \citealt{PhilPaper}). The optical off-axis afterglow represents a negligible contribution to the kilonova emission at $t<30$ days. \emph{Right column}: SED of the best-fitting model at the time of the X-ray detection 15.4 days. The best-fitting off-axis models with $E_{k}=10^{49}\,\rm{erg}$ are shown in  Fig. \ref{Fig:Off2}.}
\label{Fig:Off1}
\end{figure*}

\begin{figure*}[t!]
\center
\includegraphics[scale=0.47]{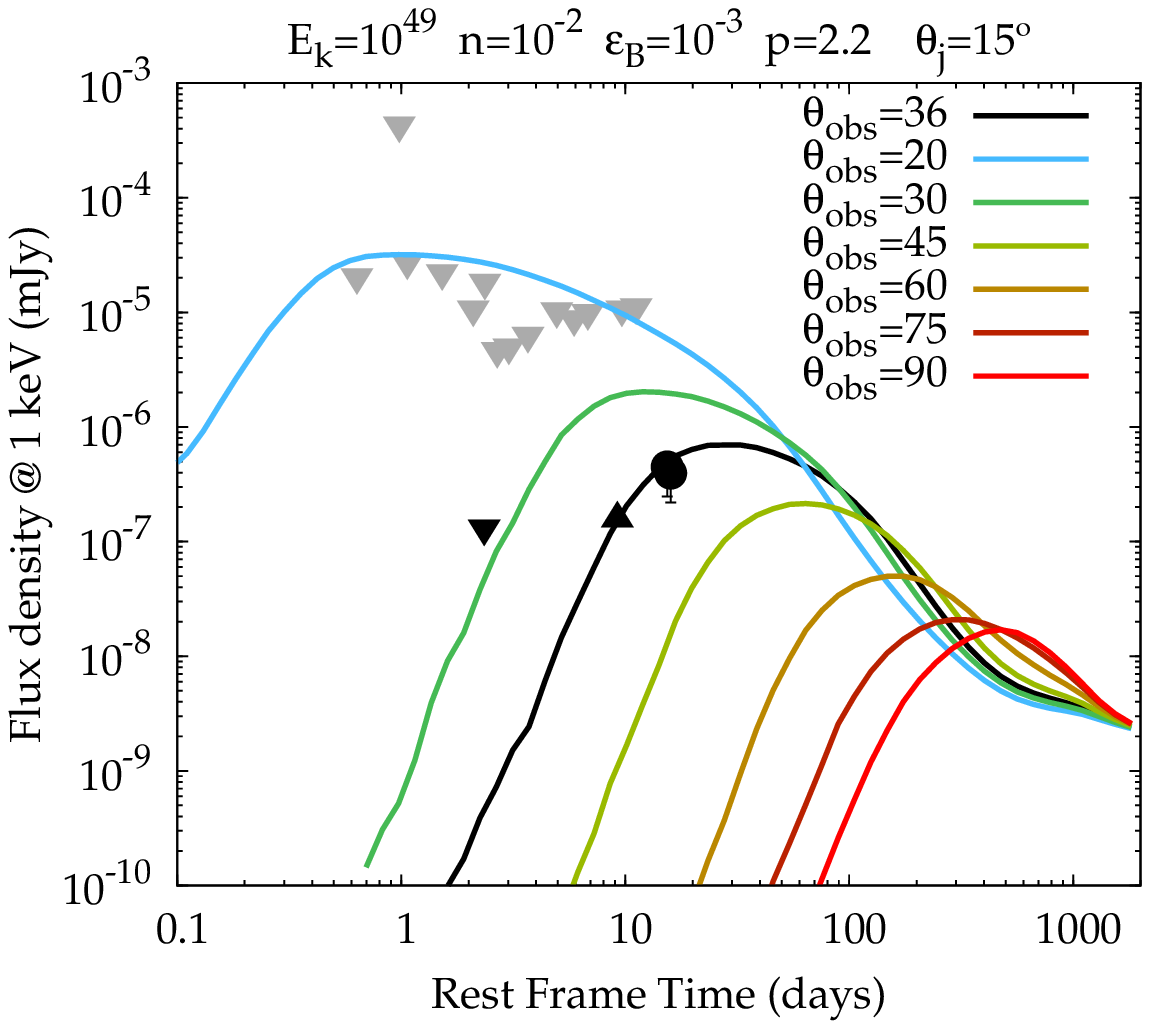}
\includegraphics[scale=0.47]{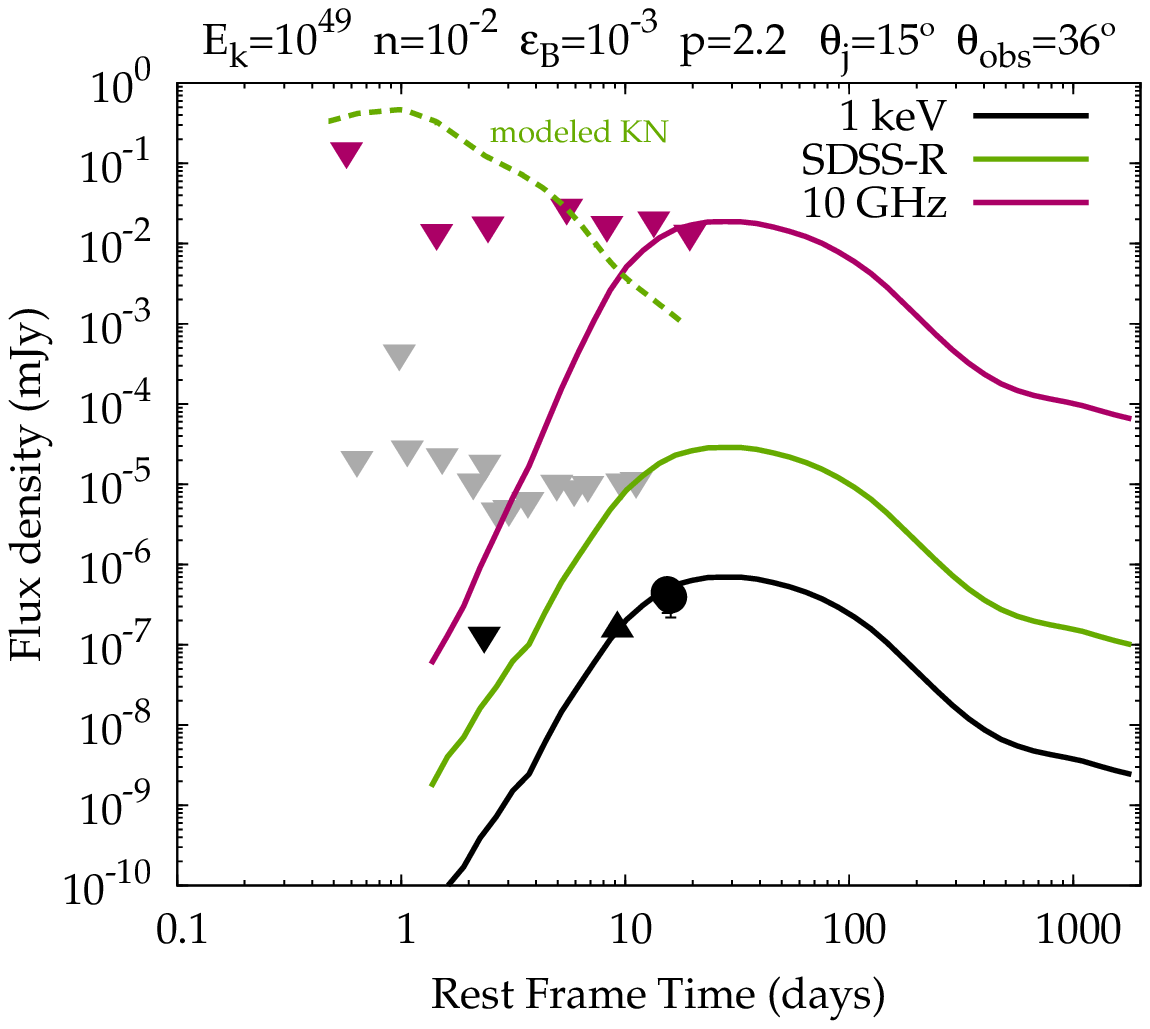}
\includegraphics[scale=0.45]{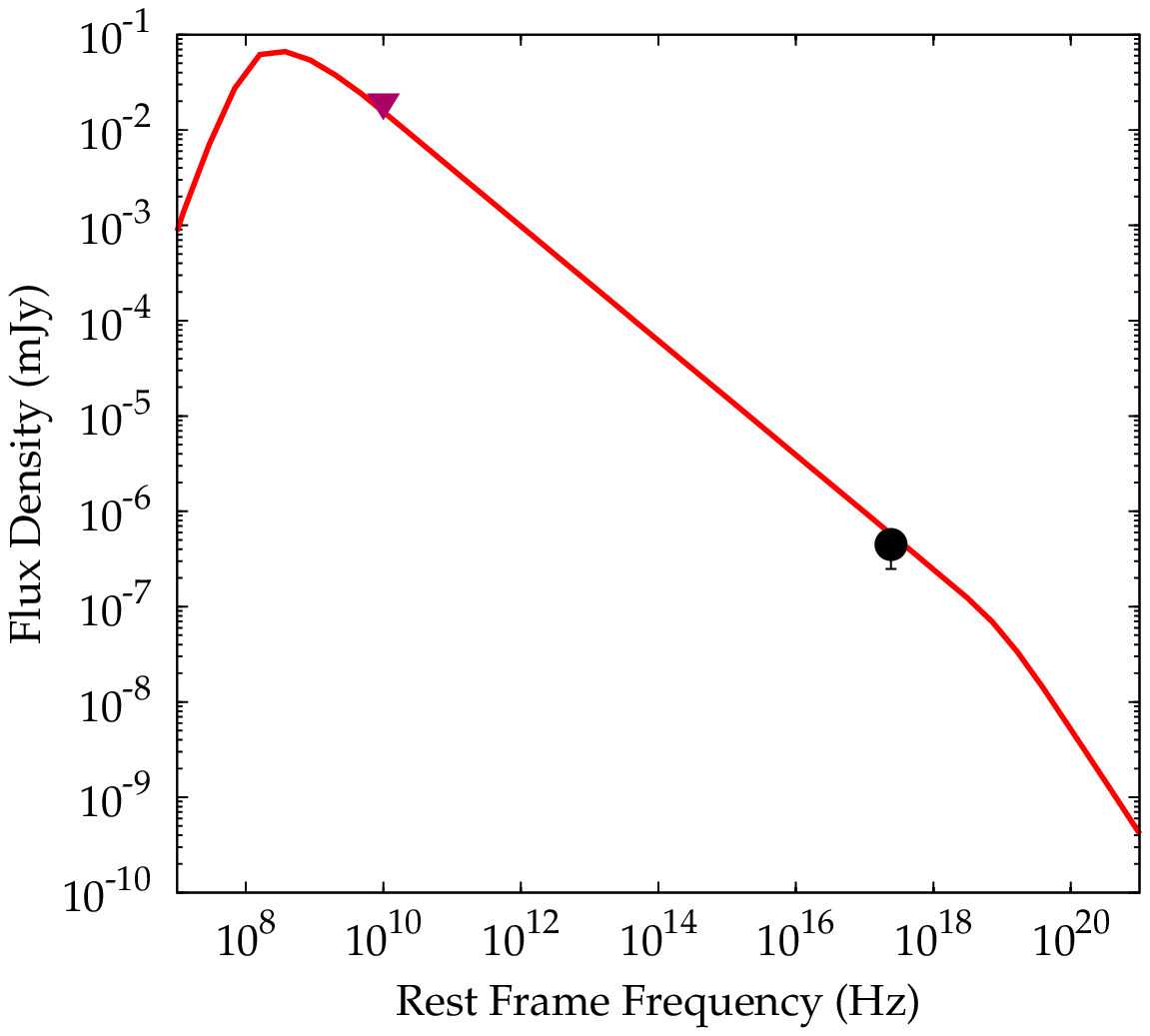}\\
\includegraphics[scale=0.47]{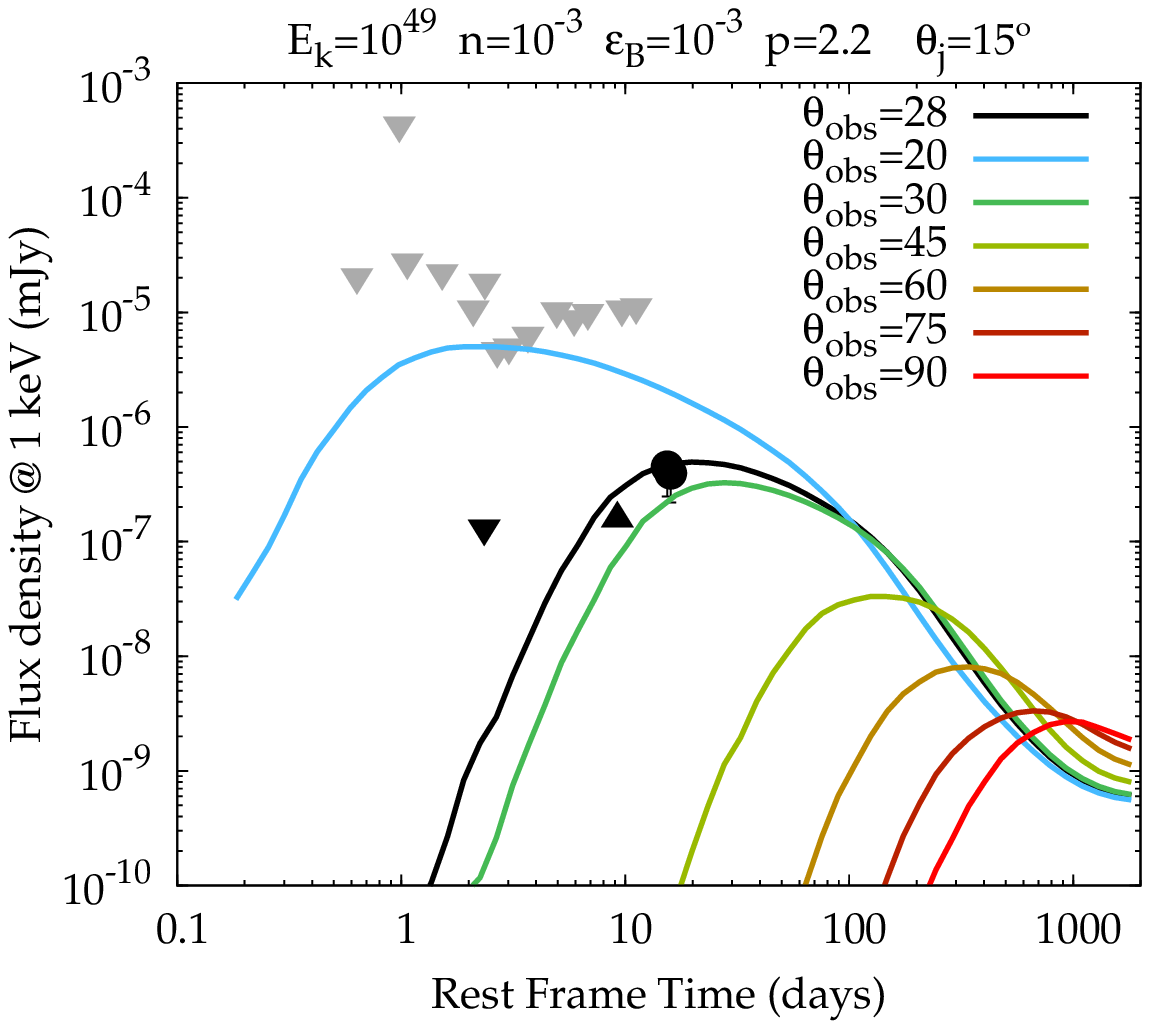}
\includegraphics[scale=0.47]{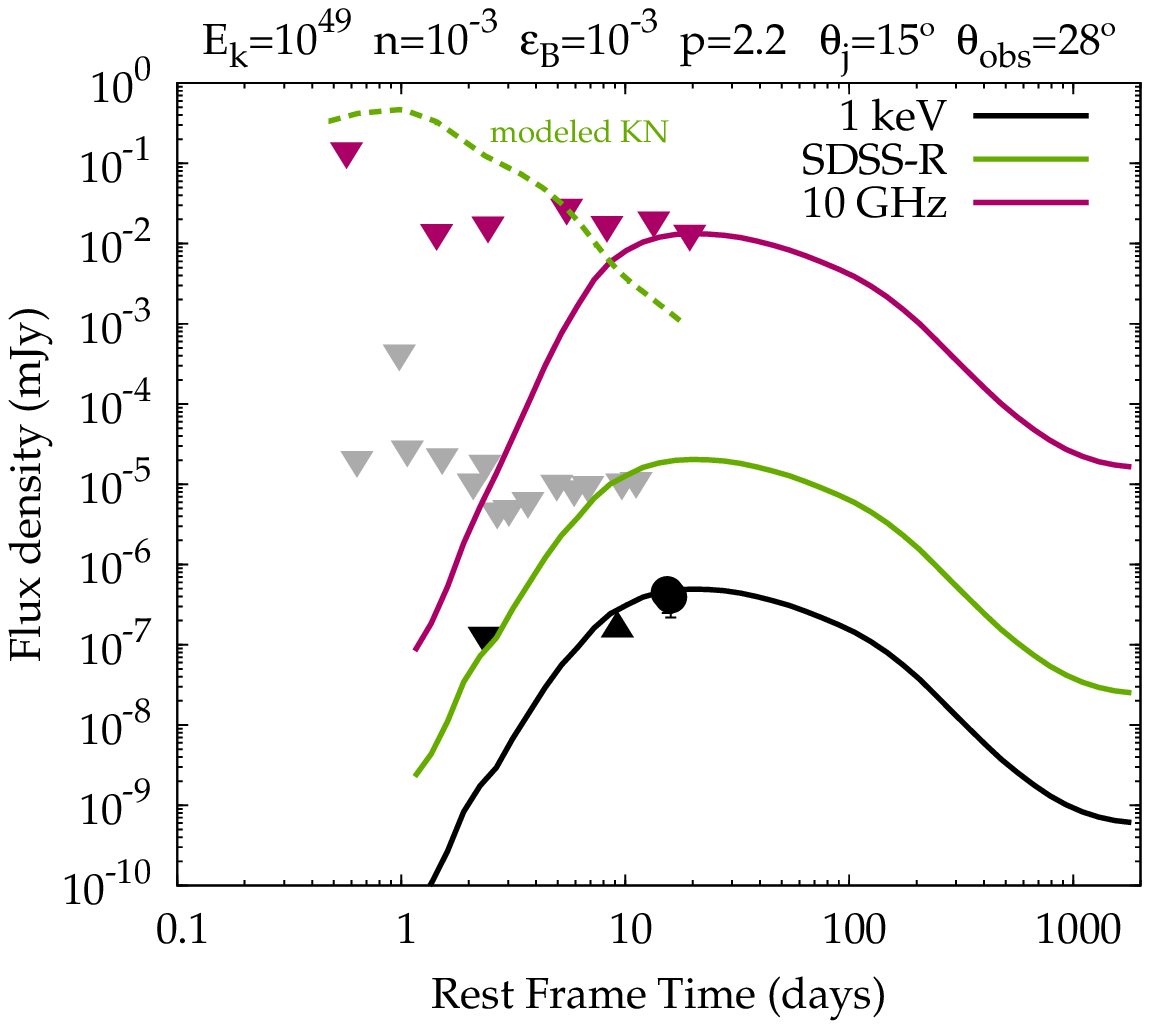}
\includegraphics[scale=0.45]{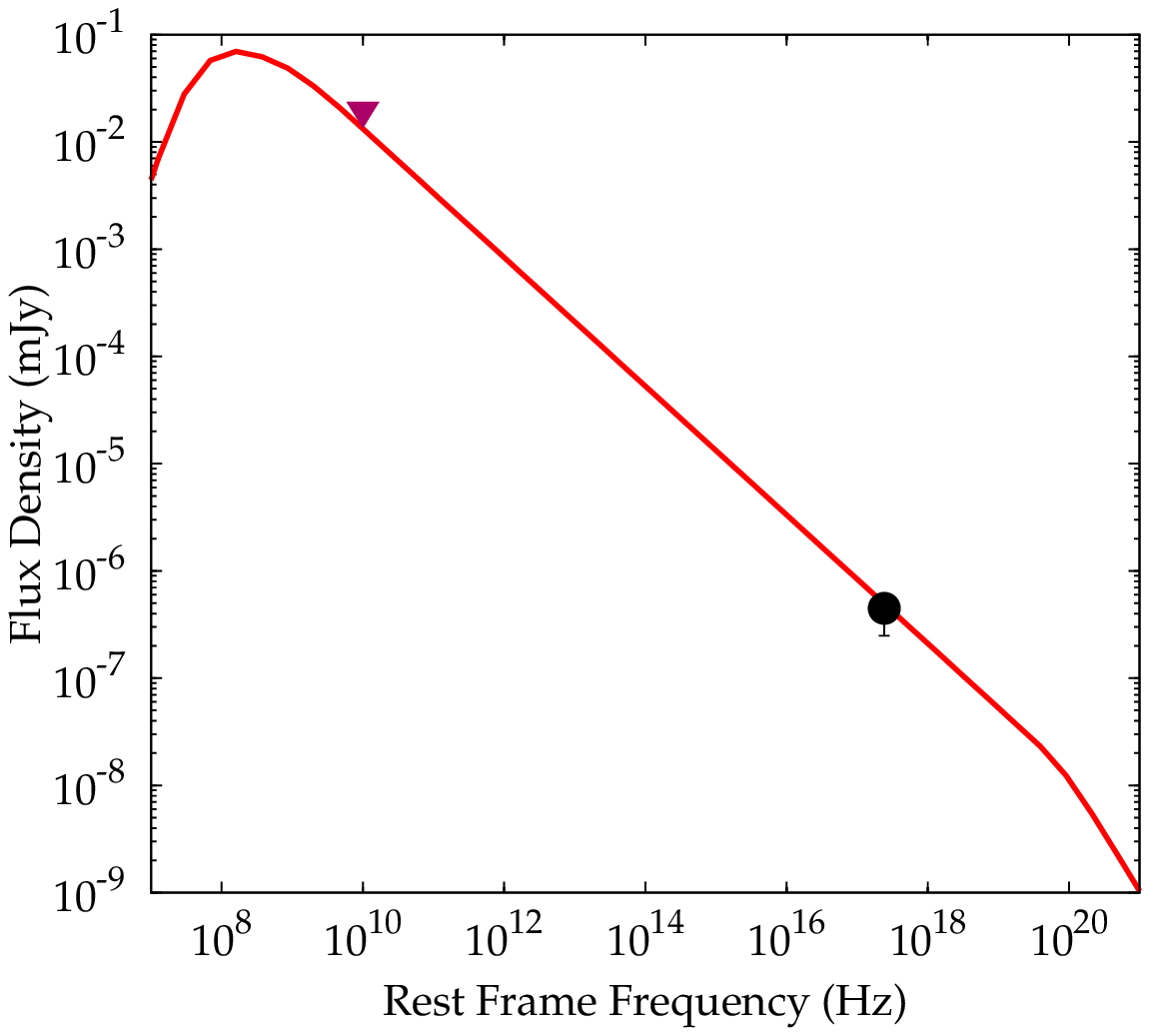}\\
\includegraphics[scale=0.47]{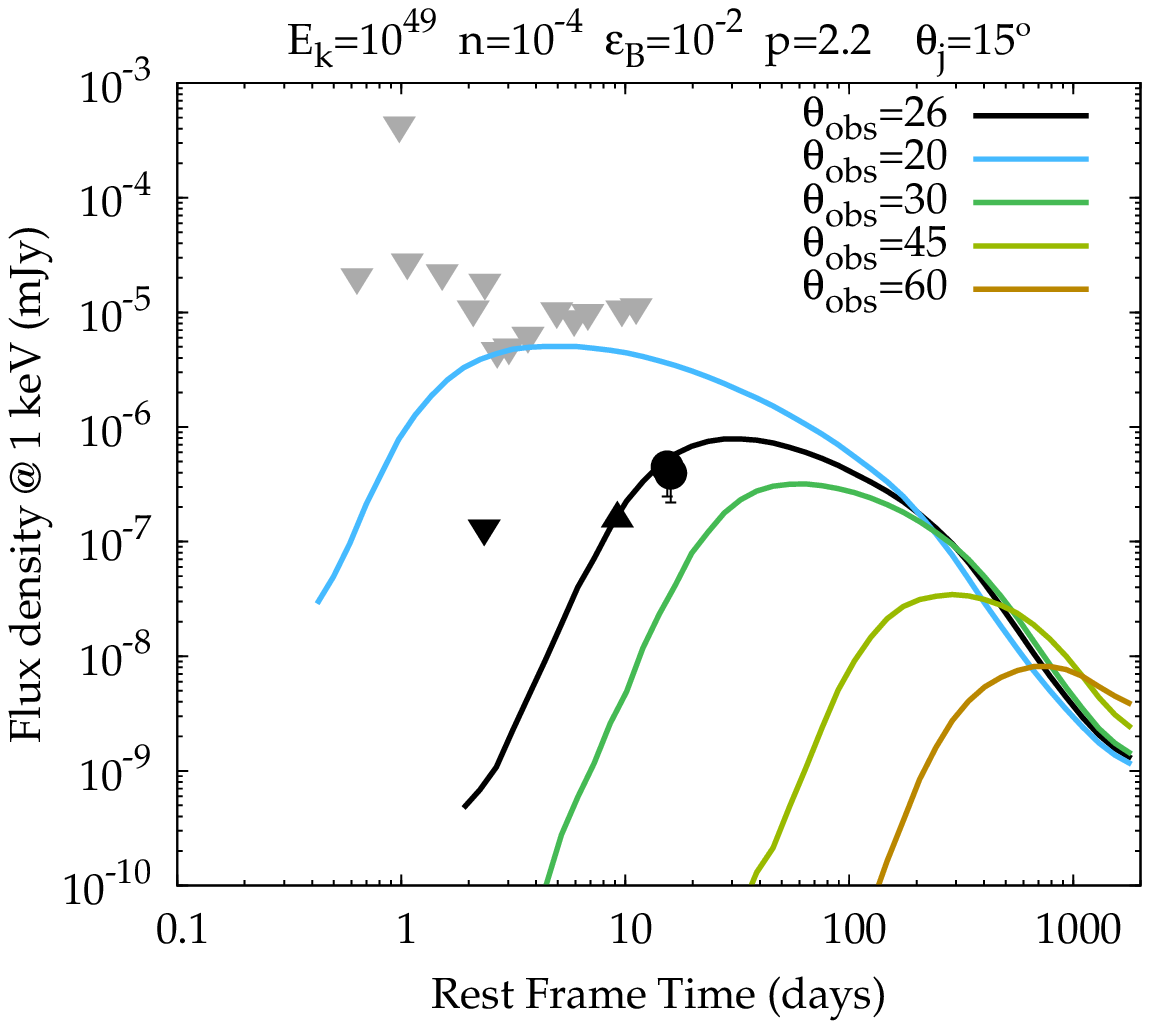}
\includegraphics[scale=0.47]{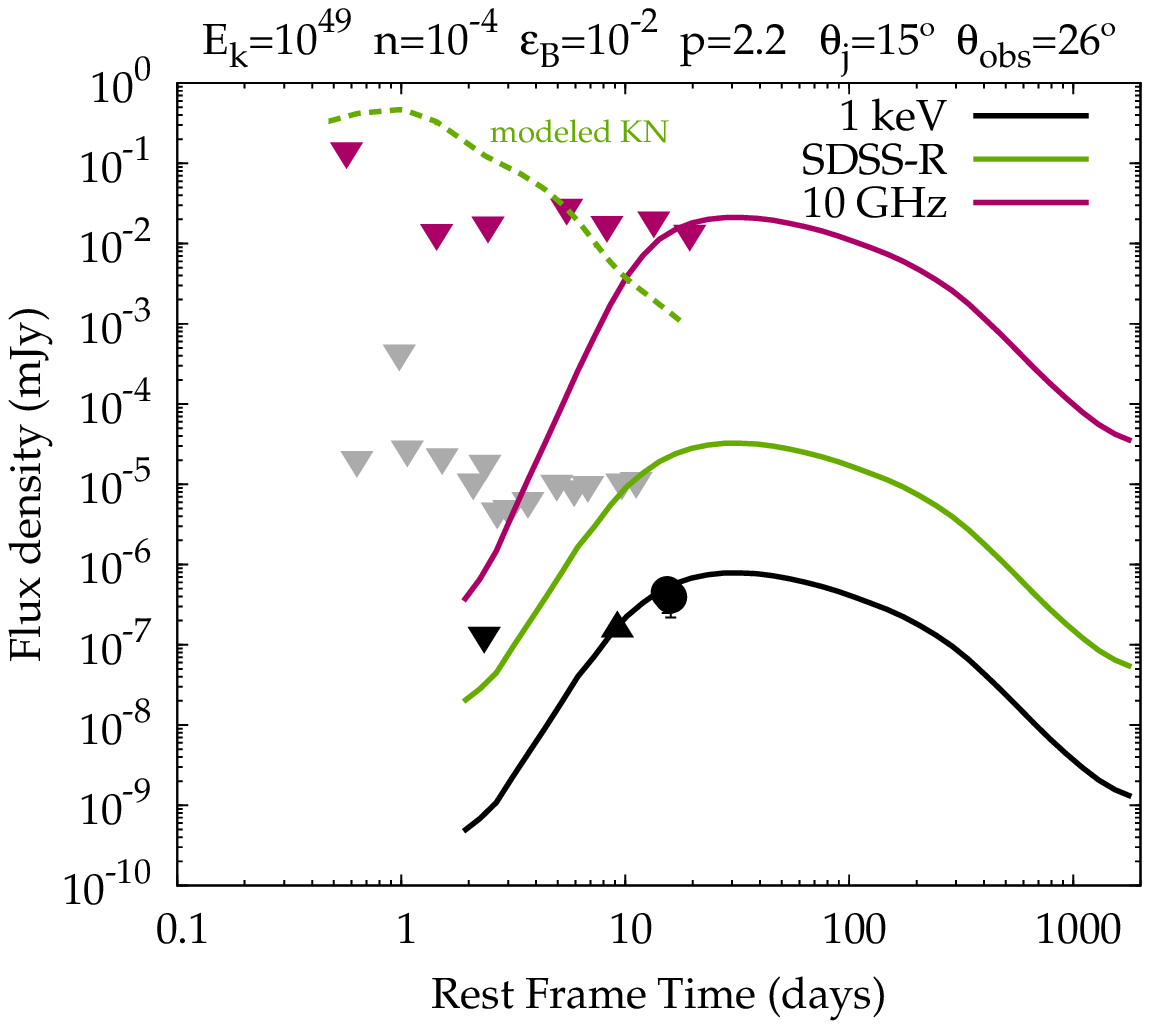}
\includegraphics[scale=0.45]{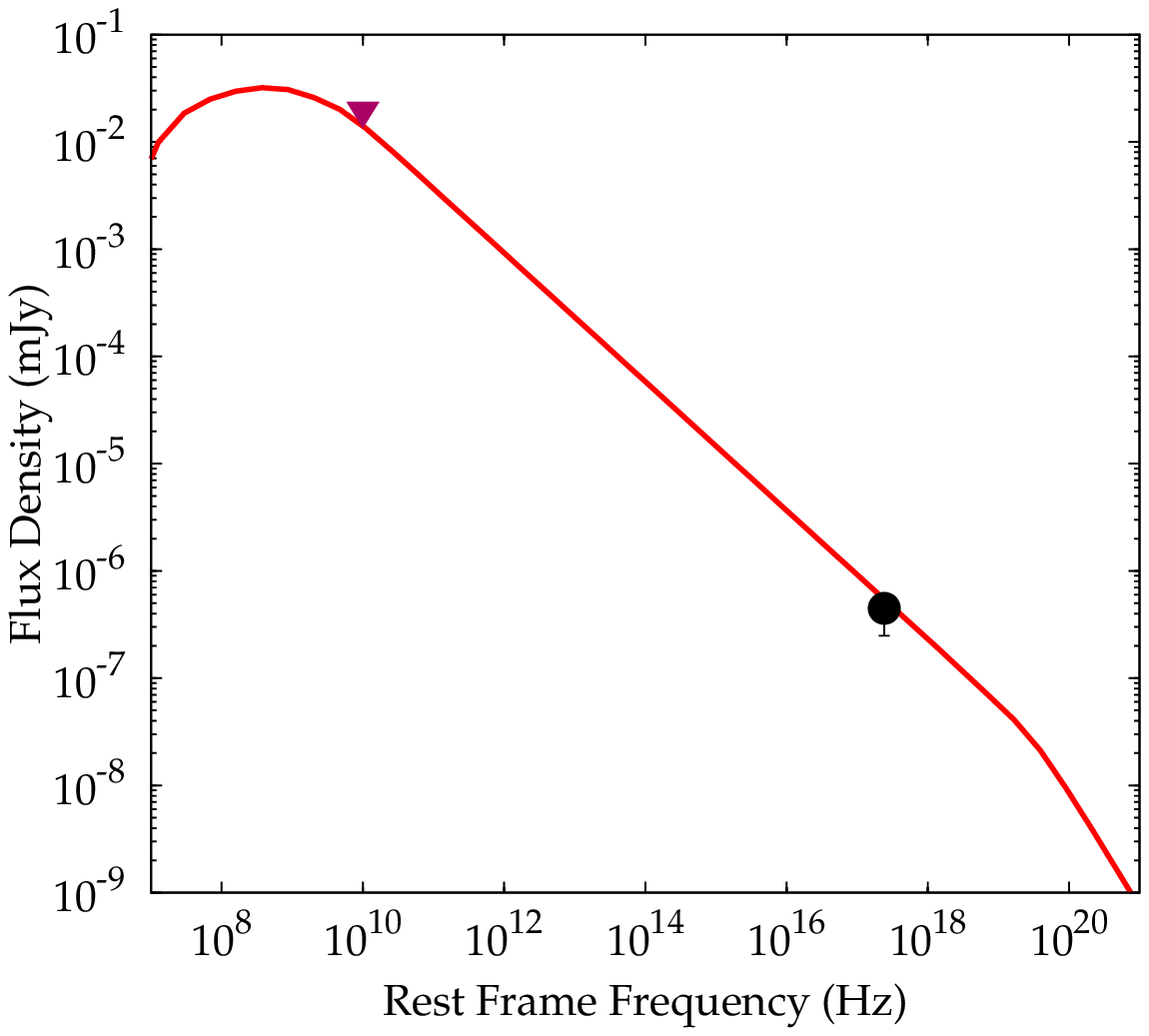}\\
\caption{Off-axis jet models with $\theta_j=15\degree$ and $E_{k}=10^{49}\,\rm{erg}$ that best represents the current set of X-ray and radio observations. Each row is dedicated to a jet-model with a given set of parameters $n$, $\epsilon_B$. Color-coding as in Fig. \ref{Fig:Off1}.  \emph{Left column}: X-ray emission for a jet with parameters indicated in each plot title and for observers at different $\theta_{obs}$ (colored lines). The black line identifies the best-fitting model. Grey triangles: Swift-XRT upper limits. Black symbols: CXO observations. \emph{Central panel}: radio (10 GHz, solid purple line) and optical emission (r-band, solid green line) for the best-fitting model compared to our VLA limits (purple triangles, \citealt{KatePaper}) and emission from the kilonova (green dashed line, \citealt{PhilPaper}). The optical off-axis afterglow represents a negligible contribution to the kilonova emission at $t<30$ days. \emph{Right column}: SED of the best-fitting model at the time of the X-ray detection 15.4 days. The best-fitting off-axis model for $E_{k}=10^{50}\,\rm{erg}$ is shown in  Fig. \ref{Fig:Off1}.}
\label{Fig:Off2}
\end{figure*}

A delayed onset of the X-ray emission can originate from the presence of an off-axis jet, originally pointed away from our line of sight.
For a simple model of a point source at an angle $\theta_{obs}$, moving at a Lorentz factor $\Gamma$, the peak in the light curve occurs when the beaming cone widens enough to engulf the line of sight, $\Gamma(t_{pk}) \sim 1/\theta_{obs}$ (e.g. \citealt{Granot02b}). This is a purely dynamical effect that does not depend on the micropysical parameters $\epsilon_e$ and $\epsilon_B$ (which instead concur to determine the overall luminosity of the emission).
From \cite{Granot02}, the evolution of the Lorentz factor of a blastwave propagating into an ISM medium can be parametrized as $\Gamma(t)\sim 6.68 (E_{k,iso,52}/n_0)^{1/8}t_{days}^{-3/8}$, which gives $\theta_{obs}\sim 0.15 (E_{k,iso,52}/n_0)^{-1/8}t_{pk,days}^{3/8}$ or $\theta_{obs}\sim 0.2\,(E_{k,50}/n_{0})^{-1/6} t_{pk,days}^{1/2}$. Before peak the off-axis model predicts a steep rise, with the flux scaling  $\propto t^{2}$. As we argued above, the mild temporal evolution of the detected X-ray emission suggests a peak not too far from our last epoch of observation at $\sim15$ days. We find $\theta_{obs}\sim(15\degree-30\degree)(E_{k,50}/n_{-3})^{-1/6}\deg$ for $t_{pk}=15-70$. If GW170817 harbored a relativistic off-axis jet with similar parameters to cosmological short GRBs ($E_{k}\sim10^{49-50}\,\rm{erg}$ and $n\sim 10^{-3}\,\rm{cm^{-3}}$, \citealt{Fong15}), this simple analytical scaling suggests off-axis angles $\theta_{obs}\sim 20\degree-40\degree$.

The actual values of the flux detected (and undetected) in the X-rays and radio pose additional constraints that break the model degeneracy in $E_k$ and $n$ as a function of $\epsilon_e$ and $\epsilon_B$. We employ realistic simulations of relativistic jets propagating into an ISM medium to fully capture the effects of lateral jet spreading with time, finite jet opening angle and transition into the non-relativistic regime.
To this aim, we run the publicly available code {\tt BOXFIT} (v2; \citealt{vanEerten10,vanEerten12}), varying $E_{k}$, $n$, $p$,  $\epsilon_B$ and $\theta_j$ (jet opening angle), and calculate the off-axis afterglow emission as observed from different lines of sight $\theta_{obs}$, with $\theta_{obs}$ varying from $5\degree$ to $90\degree$ (i.e. equatorial view). We explore a wide portion of parameter space corresponding to $E_{k}=10^{48}-10^{51}\,\rm{erg}$, $n=10^{-4}-1\,\rm{cm^{-3}}$, $\epsilon_B=10^{-4}-10^{-2}$. In our calculations we assume the fiducial value $\epsilon_e=0.1$ (e.g. \citealt{Sironi15}). For each parameter set, we consider two values for the power-law index of the electron distribution $p=2.4$ (median value from short GRBs afterglows from \citealt{Fong15}) and $p=2.2$ (as expected from particle acceleration in the ultra-relativistic limit, \citealt{Sironi15}), and we run each simulation for a collimated $\theta_j=5\,\degree$ jet and a jet with $\theta_j=15\,\degree$, representative of a less collimated outflow. As a comparison, the measured $\theta_j$ in short GRBs range between $3\degree$ and $10
\degree$ with notable lower limits $\theta_j>15\degree$ and $\theta_j>25\degree$ for GRBs\,050709 and 050724A (\citealt{Fong15} and references there in).

The results from our simulations can be summarized as follows: (i) While we find a set of solutions with $p=2.4$ that can adequately fit the X-ray light-curve, all of these simulations violate our radio limits as we detail in \cite{KatePaper}. Models with $p\geq2.4$ are ruled out and we will not discuss these simulations further.  (ii)  Models that intercept the measured X-ray flux, but with $t_{pk}\gg 15$ days, overpredict the radio emission, for which we have observations extending to $t\sim40$ days \citep{KatePaper}. Jets with $E_{k}>10^{50}\,\rm{erg}$ belong to this category and are not favored. (iii) Most high-density environments with $n\sim 0.1-1\,\rm{cm^{-3}}$ cause an earlier deceleration of the jet. As a consequence, these models require $\theta_{obs}$ between $40\degree$ and $60\degree$ to match the X-ray flux evolution (i.e. a range of $\theta_{obs}$ not favored by the early blue colors of the kilonova, \citealt{PhilPaper,MattPaper}) and are not consistent with the radio limits. (iv) Low-energy jets with $E_{k}\sim10^{48}\,\rm{erg}$ also have shorter deceleration times and require $\theta_{obs}>45\degree$ to explain the X-ray observations (and are consequently not favored by the kilonova colors). (v) Finally, wider jets have a larger allowed parameter space and are favored based on their broader light-curves around peak time.   

\begin{figure}[t!]
\center
\includegraphics[scale=0.45]{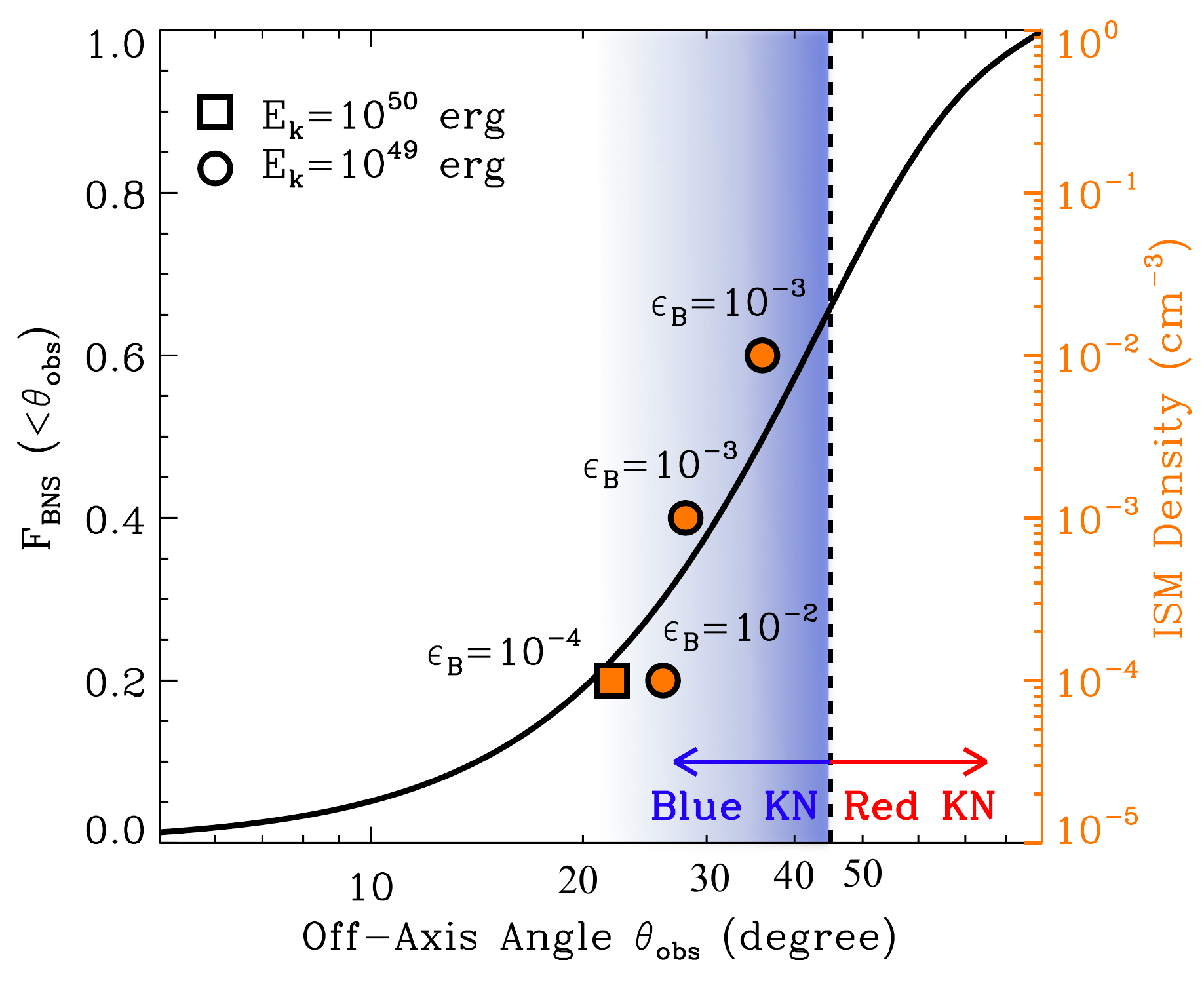}
\caption{Left y-axis and black thick line: cumulative GW detection probability at observing angles $<\theta_{obs}$ with respect to the binary axis, calculated following \cite{Metzger12} and \cite{Schutz11}. Orange points: $\theta_{obs}$ as inferred from our simulations of off-axis jets with $E_{k}=10^{49}\,\rm{erg}$ (filled circles) or $E_{k}=10^{50}\,\rm{erg}$ (filled square) and $\theta_j=15\degree$ that satisfy all the observational constraints from  X-ray and radio observations currently available, as a function of the ISM density $n$ (right y-axis). A kilonova with blue colors is expected for $\theta_{obs}\leq 45\degree$ \citep{Sekiguchi16}, shaded blue area. The value of $\epsilon_B$ for each successful simulation is also reported in the plot.}
\label{Fig:Angle}
\end{figure}

We identify a family of solutions that adequately reproduce the current data set across the spectrum (Figures \ref{Fig:Off1}-\ref{Fig:Off2}). The successful models are characterized by an off-axis jet with $10^{49}\rm{erg} \lesssim E_{k}\lesssim 10^{50}\,\rm{erg}$, $\theta_j=15\degree$ viewed $\sim 20\degree-40\degree$ off-axis and propagating into an ISM with $n\sim 10^{-4}-10^{-2}\,\rm{cm^{-3}}$, depending on the value of $\epsilon_B=10^{-4}-10^{-2}$. The dependency of the best fitting $\theta_{obs}$ values on $n$ and $\epsilon_B$ is illustrated in Fig. \ref{Fig:Angle}. The successful models are portrayed in Fig. \ref{Fig:Off1}-\ref{Fig:Off2}. Collimated outflows with $\theta_j=5\degree$ satisfy the observational constraints only for $E_{k}=10^{49}\,\rm{erg}$, $\epsilon_B=10^{-4}$, $n\sim10^{-3}\,\rm{cm^{-3}}$ and $\theta_{obs}\sim16\degree$. From Fig. \ref{Fig:Off1}-\ref{Fig:Off2} it is clear that the optical emission from the off-axis afterglow  (green line in the right-column plots) is always negligible compared to the contemporaneous kilonova emission. It is also worth noting that these models predict a radio flux density that is close to our flux limits (purple line and points), thus providing support to our tentative VLA detection at $t\sim 20$ days at the level of $\sim20\,\rm{\mu Jy}$ \citep{KatePaper}. Our favored models are not in disagreement with the radio detection of a faint transient at the level of $S/N=5$ previously reported by \cite{radio1} and \cite{radio2} $\sim$ 15 days post merger \citep{HallinanPaper}, and are fully consistent with our radio detection at 6 GHz at $t=39.4$ days, as detailed in \citep{KatePaper}.

%\tabletypesize{\small}
\begin{deluxetable}{lc}
\tablecolumns{2}
\tablewidth{0pc}
\tablecaption{{\tt BOXFIT} Parameters 
\label{tab:boxfit}}
\tablehead {
\colhead {Parameter}                &
\colhead {Values Considered}                          
}
\startdata
Jet Energy $E_{\rm k}$ (erg) & $10^{48}$, $10^{49}$, $10^{50}$, $10^{51}$ \\
Circum-merger density $n$ (cm$^{-3}$) & $10^{-4}, 10^{-3},10^{-2}, 0.1, 1$ \\
Jet opening angle $\theta_{\rm j}$ (deg) & $5, 15$ \\
Observer angle $\theta_{\rm obs}$ (deg) & $0, 5, 10, 20, 30, 45, 60, 75, 90$\\
Fraction of post-shock energy in B $\epsilon_B$& $10^{-4}$, $10^{-3}$, $10^{-2}$\\
Power-law index of electron distribution p &$2.2$, $2.4$ \\
\enddata
\tablecomments{Simulations were run at fixed values $\epsilon_e=0.1$ in a constant density medium.
 }
\end{deluxetable}

%Median parameters for cosmological short GRBs, WF paper:
%for $\epsilon_B=0.01$: $\theta_{\rm jet}=16$ (but that assumes a lot of things like cap on opening angle; $\theta_{\rm jet}=6$ for measurements only) $n=0.04$ $E_{k,iso} = 2\times10^{51}$ which is $E_k\sim 1.6\times 10^{50}$ erg.

\bigskip
%The line shows the cumulative probability for a detection at observing angle < theta with respect to the binary axis.
%I can send a ascii file with the black line, but it was easily obtained by numerically integrating equation 4 of Metzger & Berger (2012), http://iopscience.iop.org/article/10.1088/0004-637X/746/1/48/pdf , which should be given the original reference from Schutz in any paper.  

%%%%%%%%%%%%%%%%%%%%%%%%%%%%%%%%%%%%%%%%%%
\subsection{Emission from the Central Engine}

Short GRBs are sometimes accompanied by late time X-ray emission (e.g.~\citealt{Perley09,Margutti11,Fong14}), which may originate from long-lived central engine, such as an accreting black hole (e.g.~\citealt{Perna06}) or a millisecond magnetar (e.g.~\citealt{Metzger08}).

GW170817 was accompanied by luminous optical and infrared emission, consistent with predictions for the kilonova emission originating from $r$-process radioactive heating of the merger ejecta (\citealt{PhilPaper,MattPaper,RyanPaper}).  The observed X-ray transient is unlikely to originate from the central engine because the signal would be blocked by the photoelectric absorption in this same ejecta along the viewer's line of sight.  

Given the estimated ejecta mass of $\gtrsim 10^{-2}M_{\odot}$ and mean velocity $v_{\rm ej} \sim 0.1-0.2$ c (\citealt{PhilPaper,MattPaper,RyanPaper}),  the optical depth through the ejecta of radius $R \sim v_{\rm ej}t$ and density $\rho \sim M_{\rm ej}/(4\pi R^{3}/3)$ is approximately given by
\begin{eqnarray}
\tau_{\rm X} &\simeq& \rho R \kappa_{\rm X} \nonumber \\ &\approx& 90\left(\frac{\kappa_{\rm X}}{1000\rm \,cm^{2}g^{-1}}\right)\left(\frac{M_{\rm ej}}{10^{-2}M_{\odot}}\right)\left(\frac{v_{\rm ej}}{0.2\rm \, c}\right)^{-2}\left(\frac{t}{2\,\rm week}\right)^{-2}
\end{eqnarray}
where $\kappa_{\rm X} \sim1000$ cm$^{2}$ g$^{-1}$ is the expected bound-free opacity of neutral or singly-ionized heavy $r$-process nuclei at X-ray energies $\sim$ a few keV (e.g.~\citealt{Metzger17}).  The fact that $\tau_{X} \gg 1$ suggests that any X-ray signal from the engine would be highly suppressed, by a factor $e^{-\tau_{\rm X}} \ll 1$.  X-rays could escape at an earlier stage only if they were sufficiently powerful $L_{X} \gtrsim 10^{43}-10^{44}$ erg s$^{-1}$ to photo-ionize the ejecta, as is clearly not satisfied by the observed source $L_{X}\lesssim 10^{40}$ erg s$^{-1}$ \citep{MetzgerPiro14}.   

Such a high optical depth is not necessarily expected for on-axis viewers more typical of gamma-ray bursts, especially at early times when the engine is most powerful, because the relativistic jet may clear a low-density funnel through the ejecta along the binary axis.  As our orientation with GW170817 is unlikely to be so fortuitous, a central engine origin of the X-ray emission is disfavored.
%%%%%%%%%%%%%%%%%%%%%%%%%%%%%%%%%%%%%%%%%%
%\subsection{Inverse Compton}

%%%%%%%%%%%%%%%%%%%%%%%%%%%%%%%%%%%%%%%%%%
%\subsection{Thermal Emission}
%%%%%%%%%%%%%%%%%%%%%%%%%%%%%%%%%%%%%%%%%%
%ideas for other sections:
%\section{Constraints on isotropic counterparts}
%\section{Predictions for future X-ray behavior}
%%%%%%%%%%%%%%%%%%%%%%%%%%%%%%%%%%%%%%%%%%%

\section{Summary and Conclusions}
\label{Sec:Conc}
We present the first X-ray detection from a GW source thanks to CXO observations. These observations enabled the first discovery of \emph{rising} X-ray emission that we interpret in the context of isotropic or collimated outflows (on-axis and off-axis) with different properties. Our results can be summarized as follows:

\begin{itemize}
\item On-axis afterglow emission similar to that typically observed in cosmological short GRBs (i.e. $E_{k,iso}\sim 10^{51}\rm{erg}$) is clearly ruled out. 
\item A late (on-axis or isotropic) afterglow onset, due to the deceleration of a mildly relativistic outflow can explain the X-ray observations but likely violates the radio limits.
\item A central-engine origin of the X-ray emission is disfavored, as from the kilonova parameters that we infer in \cite{PhilPaper,MattPaper,RyanPaper} we derive a large optical depth that would prevent the X-rays from escaping and reach the observer. 
\item Current radio and X-ray observations are consistent with the emission from a relativistic jet with $\theta_j=15\degree$, $10^{49}\,\rm{erg}\le E_{k} \le 10^{50}\,\rm{erg}$, viewed $\sim20\degree-40\degree$ off-axis and propagating into an ISM environment with $n=10^{-4}-10^{-2}\,\rm{cm^{-3}}$ depending on $\epsilon_B=10^{-4}-10^{-2}$. Very collimated outflows with $\theta_j\sim5\degree$ are not favored by observations.
\end{itemize}

The discovery of X-ray emission from GW170817 marks a milestone in connecting on-axis GRBs with BNS mergers, and sets the stage for all future GW events with detected X-ray emission. Late-time X-ray monitoring of GW170817 at $t\ge100$ days (when it will be observable again with the CXO) will provide additional, crucial information to solve the model degeneracies and test our predictions.  Our inferences on the observing angle with respect to the jet axis might be testable using gravitational wave information from Advanced LIGO/Virgo on the binary inclination, inasmuch as the accuracy of the GW measurement is comparable to ours.

%%%%%%%%%%%%%%%%%%%%%%%%%%%%%%%%%%%%%%%%%%%
\bigskip
Support for this work was provided by the National Aeronautics and Space Administration through Chandra Award Number GO7-18024X issued by the Chandra X-ray Observatory Center, which is operated by the Smithsonian Astrophysical Observatory for and on behalf of the National Aeronautics Space Administration under contract NAS8-03060. WF acknowledges support for Program number HST-HF2-51390.001-A, provided by NASA through a grant from the Space Telescope Science Institute, which is operated by the Association of Universities for Research in Astronomy, Incorporated, under NASA contract NAS5-26555. CG acknowledges University of Ferrara for use of the local HPC facility co-funded by the ``Large-Scale Facilities 2010'' project (grant 7746/2011). Development of the Boxfit code was supported in part by NASA through grant NNX10AF62G issued through the Astrophysics Theory Program and by the NSF through grant AST-1009863. The Berger Time-Domain Group at Harvard is supported in part by the NSF through grants AST-1411763 and AST-1714498, and by NASA through grants NNX15AE50G and NNX16AC22G. DAB is supported by NSF award PHY-1707954. Simulations for BOXFITv2 have been carried out in part on the computing facilities of the Computational Center for Particle and Astrophysics of the research cooperation ``Excellence Cluster Universe'' in Garching, Germany.
%%%%%%%%%%%%%%%%%%%%%%%%%%%%%%%%%%%%%%%%%%%
\bibliographystyle{apj}
%\bibliography{margutti}

\begin{thebibliography}{}

\bibitem[\protect\citeauthoryear{{Abbott et al.}}{{Abbott et
  al.}}{2017}]{ALVdetection}
{Abbott et al.} 2017, Phys. Rev. Lett., {\it Submitted}

\bibitem[\protect\citeauthoryear{{Alexander et al.}}{{Alexander et
  al.}}{2017a}]{KatePaper}
{Alexander et al.} 2017a, \apjl, {\it Submitted}

\bibitem[\protect\citeauthoryear{{Alexander et al.}}{{Alexander et
  al.}}{2017b}]{Kategcn}
{Alexander et al.} 2017b, GRB Coordinates Network, 21897

\bibitem[\protect\citeauthoryear{{Allam et al.}}{{Allam et
  al.}}{2017}]{DECAMgcn}
{Allam et al.} 2017, GRB Coordinates Network, 21530

\bibitem[\protect\citeauthoryear{{Berger}}{{Berger}}{2014}]{Berger14}
{Berger}, E. 2014, \araa, 52, 43

\bibitem[\protect\citeauthoryear{{Blackburn et al.}}{{Blackburn et
  al.}}{2017}]{GBMgcn1}
{Blackburn et al.} 2017, GRB Coordinates Network, 21506

\bibitem[\protect\citeauthoryear{{Blanchard et al.}}{{Blanchard et
  al.}}{2017}]{PeterPaper}
{Blanchard et al.} 2017, \apjl, {\it Submitted}

\bibitem[\protect\citeauthoryear{{Blandford} \& {McKee}}{{Blandford} \&
  {McKee}}{1976}]{Blandford76}
{Blandford}, R.~D.,  \& {McKee}, C.~F. 1976, Physics of Fluids, 19, 1130

\bibitem[\protect\citeauthoryear{{Burrows} et~al.}{{Burrows}
  et~al.}{2005}]{Burrows05}
{Burrows}, D.~N., et~al. 2005, \ssr, 120, 165

\bibitem[\protect\citeauthoryear{{Cenko et al.}}{{Cenko et
  al.}}{2017}]{SWIFTgcn2}
{Cenko et al.} 2017, GRB Coordinates Network, 21572

\bibitem[\protect\citeauthoryear{{Chakraborti} et~al.}{{Chakraborti}
  et~al.}{2015}]{Chakraborti15}
{Chakraborti}, S., et~al. 2015, \apj, 805, 187

\bibitem[\protect\citeauthoryear{{Chevalier} \& {Fransson}}{{Chevalier} \&
  {Fransson}}{2006}]{Chevalier06}
{Chevalier}, R.~A.,  \& {Fransson}, C. 2006, \apj, 651, 381

\bibitem[\protect\citeauthoryear{{Chornock et al.}}{{Chornock et
  al.}}{2017}]{RyanPaper}
{Chornock et al.} 2017, \apjl, {\it Submitted}

\bibitem[\protect\citeauthoryear{{Corsi et al.}}{{Corsi et al.}}{2017}]{radio2}
{Corsi et al.} 2017, GRB Coordinates Network, 21815

\bibitem[\protect\citeauthoryear{{Coulter et al.}}{{Coulter et
  al.}}{2017a}]{SWOPEpaper}
{Coulter et al.}, {\it Submitted}

\bibitem[\protect\citeauthoryear{{Coulter et al.}}{{Coulter et
  al.}}{2017b}]{SWOPEgcn}
{Coulter et al.} 2017b, GRB Coordinates Network, 21529

\bibitem[\protect\citeauthoryear{{Cowperthwaite et al.}}{{Cowperthwaite et
  al.}}{2017}]{PhilPaper}
{Cowperthwaite et al.} 2017, \apjl, {\it Submitted}

\bibitem[\protect\citeauthoryear{{D'Avanzo} et~al.}{{D'Avanzo}
  et~al.}{2014}]{DAvanzo14}
{D'Avanzo}, P., et~al. 2014, \mnras, 442, 2342

\bibitem[\protect\citeauthoryear{{Eichler} et~al.}{{Eichler}
  et~al.}{1989}]{Eichler89}
{Eichler}, D., {Livio}, M., {Piran}, T.,  \& {Schramm}, D.~N. 1989, \nat, 340,
  126

\bibitem[\protect\citeauthoryear{{Evans et al.}}{{Evans et
  al.}}{2017a}]{SWIFTpaper}
{Evans et al.} 2017a, Science, {\it Submitted}

\bibitem[\protect\citeauthoryear{{Evans et al.}}{{Evans et
  al.}}{2017b}]{SWIFTgcn1}
{Evans et al.} 2017b, GRB Coordinates Network, 21550

\bibitem[\protect\citeauthoryear{{Fong} \& {Berger}}{{Fong} \&
  {Berger}}{2013}]{Fong13}
{Fong}, W.,  \& {Berger}, E. 2013, \apj, 776, 18

\bibitem[\protect\citeauthoryear{{Fong} et~al.}{{Fong} et~al.}{2015}]{Fong15}
{Fong}, W., {Berger}, E., {Margutti}, R.,  \& {Zauderer}, B.~A. 2015, \apj,
  815, 102

\bibitem[\protect\citeauthoryear{{Fong} et~al.}{{Fong} et~al.}{2014}]{Fong14}
{Fong}, W., et~al. 2014, \apj, 780, 118

\bibitem[\protect\citeauthoryear{{Fong et al.}}{{Fong et al.}}{2017a}]{WFpaper}
{Fong et al.} 2017a, \apjl, {\it Submitted}

\bibitem[\protect\citeauthoryear{{Fong et al.}}{{Fong et
  al.}}{2017b}]{Chandra2gcn}
{Fong et al.} 2017b, GRB Coordinates Network, 21768

\bibitem[\protect\citeauthoryear{{Gehrels} et~al.}{{Gehrels}
  et~al.}{2004}]{Gehrels04}
{Gehrels}, N., et~al. 2004, \apj, 611, 1005

\bibitem[\protect\citeauthoryear{{Goldstein et al.}}{{Goldstein et
  al.}}{2017}]{GBMdetection}
{Goldstein et al.} 2017, \apjl, {\it Submitted}

\bibitem[\protect\citeauthoryear{{Granot} et~al.}{{Granot}
  et~al.}{2002}]{Granot02b}
{Granot}, J., {Panaitescu}, A., {Kumar}, P.,  \& {Woosley}, S.~E. 2002, \apjl,
  570, L61

\bibitem[\protect\citeauthoryear{{Granot} \& {Sari}}{{Granot} \&
  {Sari}}{2002}]{Granot02}
{Granot}, J.,  \& {Sari}, R. 2002, \apj, 568, 820

\bibitem[\protect\citeauthoryear{{Haggard et al.}}{{Haggard et
  al.}}{2017a}]{HaggardPaper}
{Haggard et al.} 2017a, \apjl, {\it Submitted}

\bibitem[\protect\citeauthoryear{{Haggard et al.}}{{Haggard et
  al.}}{2017b}]{HaggardGCN1}
{Haggard et al.} 2017b, GRB Coordinates Network, 21798

\bibitem[\protect\citeauthoryear{{Hallinan et al.}}{{Hallinan et
  al.}}{2017}]{HallinanPaper}
{Hallinan et al.}, {\it Submitted}

\bibitem[\protect\citeauthoryear{{Kalberla} et~al.}{{Kalberla}
  et~al.}{2005}]{Kalberla05}
{Kalberla}, P.~M.~W., {Burton}, W.~B., {Hartmann}, D., {Arnal}, E.~M.,
  {Bajaja}, E., {Morras}, R.,  \& {P{\"o}ppel}, W.~G.~L. 2005, \aap, 440, 775

\bibitem[\protect\citeauthoryear{{LV Scientific Collaboration}}{{LV Scientific
  Collaboration}}{2017}]{ALVgcn}
{LV Scientific Collaboration}. 2017, GRB Coordinates Network, 21509

\bibitem[\protect\citeauthoryear{{Margutti} et~al.}{{Margutti}
  et~al.}{2011}]{Margutti11}
{Margutti}, R., et~al. 2011, \mnras, 417, 2144

\bibitem[\protect\citeauthoryear{{Margutti} et~al.}{{Margutti}
  et~al.}{2013}]{Margutti13}
{Margutti}, R., et~al. 2013, \mnras, 428, 729

\bibitem[\protect\citeauthoryear{{Margutti et al.}}{{Margutti et
  al.}}{2017}]{Chandra1gcn}
{Margutti et al.} 2017, GRB Coordinates Network, 21648

\bibitem[\protect\citeauthoryear{{Metzger}}{{Metzger}}{2017}]{Metzger17}
{Metzger}, B.~D. 2017, Living Reviews in Relativity, 20, 3

\bibitem[\protect\citeauthoryear{{Metzger} \& {Berger}}{{Metzger} \&
  {Berger}}{2012}]{Metzger12}
{Metzger}, B.~D.,  \& {Berger}, E. 2012, \apj, 746, 48

\bibitem[\protect\citeauthoryear{{Metzger} \& {Piro}}{{Metzger} \&
  {Piro}}{2014}]{MetzgerPiro14}
{Metzger}, B.~D.,  \& {Piro}, A.~L. 2014, \mnras, 439, 3916

\bibitem[\protect\citeauthoryear{{Metzger}, {Quataert}, \&
  {Thompson}}{{Metzger} et~al.}{2008}]{Metzger08}
{Metzger}, B.~D., {Quataert}, E.,  \& {Thompson}, T.~A. 2008, \mnras, 385, 1455

\bibitem[\protect\citeauthoryear{{Mooley et al.}}{{Mooley et
  al.}}{2017}]{radio1}
{Mooley et al.} 2017, GRB Coordinates Network, 21814

\bibitem[\protect\citeauthoryear{{Narayan}, {Paczynski}, \& {Piran}}{{Narayan}
  et~al.}{1992}]{Narayan92}
{Narayan}, R., {Paczynski}, B.,  \& {Piran}, T. 1992, \apjl, 395, L83

\bibitem[\protect\citeauthoryear{{Nicholl et al.}}{{Nicholl et
  al.}}{2017}]{MattPaper}
{Nicholl et al.} 2017, \apjl, {\it Submitted}

\bibitem[\protect\citeauthoryear{{Perley} et~al.}{{Perley}
  et~al.}{2009}]{Perley09}
{Perley}, D.~A., et~al. 2009, \apj, 696, 1871

\bibitem[\protect\citeauthoryear{{Perna}, {Armitage}, \& {Zhang}}{{Perna}
  et~al.}{2006}]{Perna06}
{Perna}, R., {Armitage}, P.~J.,  \& {Zhang}, B. 2006, \apjl, 636, L29

\bibitem[\protect\citeauthoryear{{Piran}}{{Piran}}{2004}]{Piran04}
{Piran}, T. 2004, Reviews of Modern Physics, 76, 1143

\bibitem[\protect\citeauthoryear{{Rybicki} \& {Lightman}}{{Rybicki} \&
  {Lightman}}{1979}]{Rybicki79}
{Rybicki}, G.~B.,  \& {Lightman}, A.~P. 1979, {Radiative processes in
  astrophysics}

\bibitem[\protect\citeauthoryear{{Sari} \& {Piran}}{{Sari} \&
  {Piran}}{1999}]{Sari99b}
{Sari}, R.,  \& {Piran}, T. 1999, \apj, 520, 641

\bibitem[\protect\citeauthoryear{{Savchenko et al.}}{{Savchenko et
  al.}}{2017}]{INTEGRALgcn}
{Savchenko et al.} 2017, GRB Coordinates Network, 21507

\bibitem[\protect\citeauthoryear{{Schutz}}{{Schutz}}{2011}]{Schutz11}
{Schutz}, B.~F. 2011, Classical and Quantum Gravity, 28, 125023

\bibitem[\protect\citeauthoryear{{Sekiguchi} et~al.}{{Sekiguchi}
  et~al.}{2016}]{Sekiguchi16}
{Sekiguchi}, Y., {Kiuchi}, K., {Kyutoku}, K., {Shibata}, M.,  \& {Taniguchi},
  K. 2016, \prd, 93, 124046

\bibitem[\protect\citeauthoryear{{Sironi}, {Keshet}, \& {Lemoine}}{{Sironi}
  et~al.}{2015}]{Sironi15}
{Sironi}, L., {Keshet}, U.,  \& {Lemoine}, M. 2015, \ssr, 191, 519

\bibitem[\protect\citeauthoryear{{Soares-Santos et al.}}{{Soares-Santos et
  al.}}{2017}]{DECamPaper1}
{Soares-Santos et al.} 2017, \apjl, {\it Submitted}

\bibitem[\protect\citeauthoryear{{Soderberg} et~al.}{{Soderberg}
  et~al.}{2010}]{Soderberg10}
{Soderberg}, A.~M., et~al. 2010, \nat, 463, 513

\bibitem[\protect\citeauthoryear{{Troja et al.}}{{Troja et
  al.}}{2017}]{TrojaGCN}
{Troja et al.} 2017, GRB Coordinates Network, 21765

\bibitem[\protect\citeauthoryear{{Valenti et al.}}{{Valenti et
  al.}}{2017}]{ValentiPaper}
{Valenti et al.}, {\it Submitted}

\bibitem[\protect\citeauthoryear{{van Eerten}, {Zhang}, \& {MacFadyen}}{{van
  Eerten} et~al.}{2010}]{vanEerten10}
{van Eerten}, H., {Zhang}, W.,  \& {MacFadyen}, A. 2010, \apj, 722, 235

\bibitem[\protect\citeauthoryear{{van Eerten} \& {MacFadyen}}{{van Eerten} \&
  {MacFadyen}}{2012}]{vanEerten12}
{van Eerten}, H.~J.,  \& {MacFadyen}, A.~I. 2012, \apj, 751, 155

\bibitem[\protect\citeauthoryear{{Yang et al.}}{{Yang et al.}}{2017}]{DLT40gcn}
{Yang et al.} 2017, GRB Coordinates Network, 21531

\end{thebibliography}

\end{document}